\documentclass[aps,
english,preprintnumbers,nofootinbib,
twocolumn]{revtex4-1}

\usepackage{amsfonts,amsmath,amssymb}
\usepackage{graphicx}
\usepackage[utf8]{inputenc}
\usepackage{hyperref}
\usepackage{babel}

\usepackage{xcolor}

\newcommand\be{\begin{equation}}
\newcommand\ee{\end{equation}}
\newcommand\bea{\begin{eqnarray}}
\newcommand\eea{\end{eqnarray}}

\begin{document}

\title{ Thermalization in Krylov Basis }

\author{Mohsen Alishahiha${}^a$\email{alishah@ipm.ir} and,  Mohammad Javad Vasli${}^{b,a}$
\email{vasli@}}
\affiliation{${}^a$ School of Physics, Institute for Research in Fundamental Sciences (IPM),\\
	P.O. Box 19395-5531, Tehran, Iran\\ ${}^b$  Department of Physics, University of Guilan, P.O. Box 41335-1914, Rasht, Iran\\
 E-mails: {\tt  alishah@ipm.ir, vasli@phd.guilan.ac.ir
}}

\begin{abstract}

We study thermalization in closed non-integrable quantum systems using the Krylov basis. We demonstrate that for thermalization to occur, the matrix representation of  
typical local operators in the Krylov basis should
exhibit a specific tridiagonal form with all other
elements in the matrix are exponentially small,
reminiscent of the eigenstate thermalization hypothesis.
Within this framework, we propose that the nature of thermalization, whether weak or strong, can be examined  by the infinite time average of the Krylov complexity. Moreover, we analyze the variance of Lanczos coefficients as another probe for  the nature of thermalization. One observes that although the variance of Lanczos coefficients may capture certain 
features of thermalization, 
it is not as effective as the infinite time average of complexity.

\end{abstract}

\maketitle

\section{Introduction}

Based on our everyday experience, the thermalization of macroscopic systems is one of the most natural phenomena in nature. Although to see a macroscopic system is approaching thermal equilibrium one does not need to produce several copies of 
the system, the statistical mechanics
in which we are dealing  with ``ensembles'' 
is provided a powerful tool to study thermalization. This has to do 
with the  ergodic property of classical chaotic systems 
that validates
the statistical mechanics. In fact in these systems
 the ensemble averages  used in statistical mechanics calculations agree with the
 time averages involving in our experiments.
 

Even though  for closed quantum systems one may also
observe emerging of the thermal equilibrium in non-equilibrated
systems (caused by {\it e.g.} global quench), 
unlike the classical systems,  the thermalization may be seen without performing any time averages \cite{{Srednicki:1994mfb},{Deutsch:1991}}.  Indeed, out of equilibrium states 
approach to their thermal expectations shortly after relaxation.  It is, however, important to note that in
closed quantum systems  dynamics is unitary and time reversal invariant, 
and therefore, a priori, it
is not obvious how and in what sense the thermal equilibrium can be reached dynamically. 

The notion of thermalization in quantum mechanics may
be described by the eigenstate thermalization hypothesis
(ETH) \cite{{Srednicki:1994mfb},{Deutsch:1991}} which gives an understanding of how an observable thermalizes to its thermal equilibrium value.  According to ETH for sufficiently complex quantum systems the
energy eigenstates are indistinguishable from thermal states with the same average
energy.

Although,  it is believed that a non-integrable model will generally thermalize, the 
nature of thermalization might differ in different situations.  Actually,
besides Hamiltonian which gives dynamics of the system,  the nature of 
the thermalization may also depend on the initial state, such that, within a fixed model different initial states may exhibit different behaviors \cite{Banuls:2011vuw}. 

To explore this point better let us consider spin$-\frac{1}{2}$ Ising model given by 
the following Hamiltonian
\be\label{Ising}
H=-J\sum_{i=1}^{N-1}\sigma^z_i\sigma^z_{i+1}-\sum_{i=1}^N
(g \sigma^x_i+h\sigma^z_i)\,.
\ee
Here and in what follows $\sigma^{z,y,z}$ are Pauli matrices
and $J, g$ and $h$ are constants which define the model. By rescaling one may set
$J=1$, and the nature of the model, being chaotic or integrable, is
controlled by constants $g$ and $h$. In particular,  
for $gh\neq 0$ the
model is non-integrable. In what follows to perform our numerical computations we will set
$h=0.5,\, g=-1.05$ \cite{Banuls:2011vuw}. It has been shown in  \cite{Banuls:2011vuw}
that three different initial states in which all spins are aligned  on $x, y$ or $z$ directions denoting by
$|X+\rangle, |Y+\rangle, |Z+\rangle$ respectively, results in three distinct thermalization behaviors.  

 In general, we would  like to study time
evolution of expectation value of a local operator (observable) ${\cal O}$
\be
\langle \psi(t)| {\cal O} |\psi(t)\rangle=\rm{Tr}\left( e^{-iHt}\rho_0 e^{iHt} {\cal O}\right),
\ee
whose behavior could explore the nature of thermalization 
whether it is 
strong or weak. In the strong thermalization, the expectation  value relaxes to the thermal value very fast, while  for weak thermalization  it strongly oscillates around the thermal value, though its
time average attains the thermal value. Here  $\rho_0$ is density 
state associated with the initial state $|\psi_0\rangle$.

For the Ising model \eqref{Ising} it has been shown  that although  the initial state $|Y+\rangle$ exhibits strong thermalization, for initial state $|Z+\rangle$ 
one observes weak thermalization and 
for initial state $|X+\rangle$ there is an 
apparent departure of the thermal expectation value 
form its thermal value suggesting that there might be no thermalization for this state\footnote{ Actually it seems that the apparent departure of thermalization in this case is due to the finite $N$ effects and indeed, even in this case we still have a weak thermalization \cite{Sun:2020ybj}.} \cite{Banuls:2011vuw}.

It was proposed in \cite{Banuls:2011vuw} that whether we are going to observe strong or weak thermalization is
closely related to the effective inverse temperature, $\beta$, of the 
initial state which can be read from the following equation 
\be\label{beta0}
{\rm Tr}\left(H(\rho_0-\rho_{th})\right)=0,
\ee
where $\rho_{th}=
\frac{ e^{-\beta H}}{{\rm Tr}(e^{-\beta H})}$ is thermal density state with inverse temperature $\beta$. The strong thermalization 
occurs when the effective inverse temperature
of initial states is close to zero. On the other hand, for initial states whose effective inverse temperature 
are sufficiently
far away from zero, one observes weak thermalization. In particular, for
the model given in  \eqref{Ising} the
 initial state $|Y+\rangle$  has zero effective
inverse temperature and  for initial states $|Z+\rangle$ and $|X+\rangle$  one has $\beta= 0.7275$ and  $\beta= -0.7180$, respectively.

For a given initial state $|\psi_0\rangle$ the equation \eqref{beta0} may be rewritten as follows
\be
{\rm Tr}\left(\rho_{th}H)\right)=\langle\psi_0 |H|\psi_0\rangle=E\,,
\ee
which suggests that the information of the effective inverse temperature could be
read from the expectation value of the energy.
Indeed,  the regime  on which the strong or weak thermalization may occur 
could also be identified by the normalized energy of the initial state \cite{Chen:2021}
\be\label{NE}
{\cal E}=\frac{\langle \psi_0 |H|\psi_0\rangle-E_{min}}{E_{max}-E_{min}}
\ee
where $E_{max}, E_{min}$ are maximum and minimum energy eigenvalues of the Hamiltonian.
Actually, the quasiparticle explanation of weak thermalization suggests that initial states with weak thermalization are in the regime of near the edge of 
energy spectrum \cite{Lin:2016egw}.

Although in the literature,  mainly, the normalized energy \eqref{NE} has been
considered to study weak and strong thermalization, it is found useful to work with 
the expectation value of energy itself which contains the same amount of information as that of the normalized energy. 
  
 To further explore the nature of thermalization in the Ising model 
 \eqref{Ising}, let us consider an arbitrary initial state in the Bloch sphere which may
 be parameterized by two angles $\theta$ and $\phi$ as follows\footnote{In general the initial state could be identified by
$2N$ angles  $(\theta_i,\phi_i)$ for $i=1,\cdots N$. In our case we have 
assumed that angles in all sites are equal. }
\be\label{initial}
|\theta,\phi\rangle =\prod_{i=1}^{N}\left(\cos\frac{\theta}{2}\;\; |Z+\rangle_i+e^{i\phi}\sin\frac{\theta}{2}\;\; |Z-\rangle_i\right)\,,
\ee
where $|Z\pm\rangle $ are eigenvectors of $\sigma^z$ with eigenvalues $\pm$. Indeed, at each site, the  corresponding state is the eigenvector of the operator ${\cal O}_i=n\cdot \sigma_i$, with 
$n$ is the unit vector on the Bloch sphere. More explicitly, one has
\be\label{OPi}
{\cal O}_i(\theta,\phi)=n\cdot \sigma_i=\cos\theta\,\; \sigma^z_i+\sin\theta\,(\cos\phi\, \sigma^x_i+
\sin\phi\,\; \sigma^y_i),
\ee
for $i=1,\cdots, N$.

For this general initial state and for the model \eqref{Ising}
 one can compute the expectation value of energy
 which has the following simple form 
\be\label{EE}
E=-\cos\theta\left(Nh+(N-1)J\cos\theta\right)
-Ng\cos\phi\,\sin\theta\,.
\ee
An interesting feature of 
the expectation value of energy is that for large $N$ ($N\gg 1$),
the number of spins appears as an overall factor and thus the density 
of energy defined by $E/N$  (energy per site) is independent of the size of the 
system
\be\label{ED}
\frac{E}{N}\approx-\left(h\cos\theta+J\cos^2\theta\right)
-g\cos\phi\,\sin\theta,\,\,\,\, N\gg 1 .
\ee

\begin{figure}[h!]
	\begin{center}
			\includegraphics[width=0.8\linewidth]{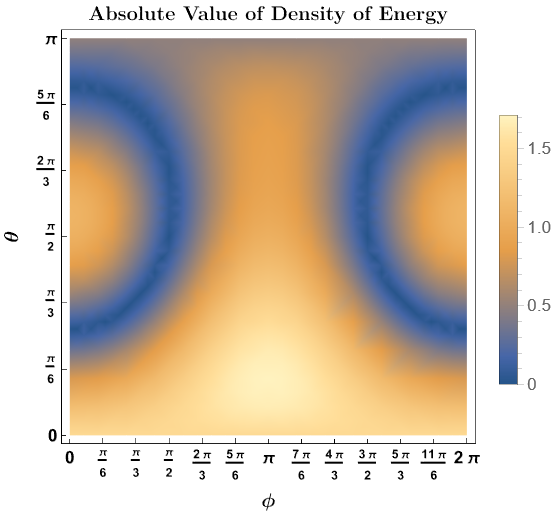}
  	\end{center}
	\caption{Absolute value of the density of energy evaluated using the
	analytic expression \eqref{EE} for $N=100$ and $J=1,\,h=0.5,\, g=-1.05$. 
}
	\label{fig:Ising1}
\end{figure}

Using this analytic expression for the expectation value of energy we have drawn the density of energy in 
 figure \ref{fig:Ising1} for $N=100$ and $J=1,\,h=0.5,\,g=-1.05$. Actually, to highlight the regions where
 the density of energy vanishes we have depicted its absolute value.

\begin{figure}[h!]
	\begin{center}
		\includegraphics[width=0.8\linewidth]{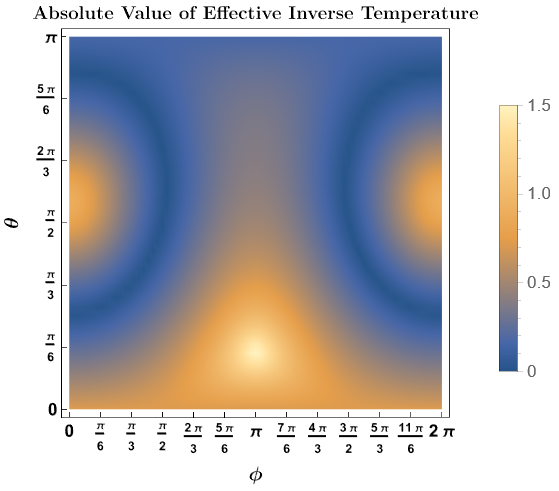}
  	\end{center}
	\caption{Absolute value of  the effective inverse temperature for 
 arbitrary $\theta, \phi$ for the general  initial state \eqref{initial}.  Here we have set $N=7$ and $g=0.5,\; h=-1.05$. 
}
	\label{fig:Ising2}
\end{figure}

To compare this result with the behavior of the effective inverse 
temperature and in particular the sensitivity of the result with the size of the system,
in figure \ref{fig:Ising2} we have presented the numerical result of the absolute value 
of the effective inverse temperature for $N=7$. Since the Hamiltonian of the model \eqref{Ising} is traceless, the locus of  $\beta=0$
are given by the regions over which $E=0$ that are shown by 
two dark semi circles (ring of zero $\beta$\footnote{By a phase shift one may draw the density of energy for $-\pi \leq \phi
\leq \pi$ for which $\beta=0$ region is a ring.}) in figures \ref{fig:Ising1} and \ref{fig:Ising2}.
This is, particularly, illustrative since the main significant information contained in  $\beta$ is its distance (absolute value) from zero. Generally, it is believed that strong thermalization occurs near the ring of zero $\beta$.

One observes that the behavior of the effective inverse temperature matches 
 exactly that of the density of energy event though the size of the two systems by which 
these quantities are evaluated are different by about a factor of 15. This shows 
the robustness of the results against the size of the system. 
In particular, this has to be 
compared with the results in the literature where the numerical computations
have been performed for $N=14$.  Even though our $\beta$ is 
evaluated for $N=7$ in comparison with that of $N=14$ the error we 
 acquire is about ${\cal O}(1)$  percent.

This article aims to study quantum thermalization using the Krylov basis which seems to be a more appropriate basis when the dynamics of the system is our interest.  Although the Krylov method has been used to study numerical computations \cite{viswanath1994recursion}, in recent years there have been several activities to use the Krylov method in the context of quantum chaos (see \cite{Parker:2018yvk} and its citations). 
 
In this paper, we would like to explore a potential application of Krylov space within the context of quantum thermalization. The key advantage of studying thermalization in this framework lies in the fact that, under a unitary time evolution the trajectory of a given initial state does not necessarily expose into
the entire Hilbert space. Instead, it remains confined within a subset known as the Krylov space, which typically has a smaller dimension compared to the full Hilbert space of the system. Thus, focusing on the Krylov space suffices for studying the time evolution of the system. In particular, if 
the system has conserved charges, working in this basis we are 
automatically confined in a subsystem that preserves the symmetry of the initial state. 

The paper is organized as follows. In the next section, we will study the late time behavior of the expectation value of typical operators within the Krylov basis. Following the Eigenstate Thermalization Hypothesis (ETH), we will propose an ansatz for the matrix elements of the operator within the Krylov basis, called the Krylov Thermalization Hypothesis (KTH). We will present several numerical computations in support of our ansatz. In section three we will study the nature of thermalization in this framework. 
Specifically, we will introduce two metrics to probe the 
nature of thermalization—namely, the variance of Lanczos 
coefficients and the infinite time average of Krylov 
complexity. While the former may not provide a definitive 
conclusion, the latter offers a pattern  that agrees 
perfectly with other proposed probes in the existing literature.
Additionally, we will calculate the inverse participation 
ratio to contrast it with the results from the infinite time 
average of complexity. 
The late section is devoted to discussions.


\section{Krylov basis and thermalization }

Let us consider a  closed quantum system with time independent local Hamiltonian $H$ whose eigenstates and eigenvalues 
are denoted by $|E_n\rangle$, and $E_n$, respectively. Starting with an initial state, 
$|\psi_0\rangle$, in the Schr\"odinger picture 
at any time one has
\be\label{Schr}
|\psi(t)\rangle=e^{iHt}|\psi_0\rangle\,.
\ee 
In the  context of quantum thermalization  the main purpose 
is to start with an initial state and then quickly alter the system, 
{\it e.g.} by 
a global quench,  and then let the system evolve under the local Hamiltonian $H$. 
As we have already mentioned, generally,  we are interested in the late time behavior of the expectation value 
of  local operators (observables)  
 \be\label{EV}
\langle \psi(t)| {\cal O} |\psi(t)\rangle=\langle\psi_0| e^{-iHt}{\cal O} e^{iHt}|\psi_0 \rangle=\langle {\cal O}(t)\rangle\,.
\ee
 The main  question
is to what extent and for what times the system
can be described by a suitable thermal equilibrium system in which the above   expectation value 
can be approximated by  ${\rm Tr}\left(\rho_{th}{\cal O}\right)$. 

To study different features of chaotic systems and thermalization one usually utilizes 
the energy spectrum and energy eigenstates which amounts to diagonalize the Hamiltonian. For example, the
nature of quantum chaos may be given in terms of the energy level statistics \cite{Bohigas:1983er}. 

In the energy eigenstates, assuming $|\psi_0\rangle=\sum_\alpha c_\alpha|E_\alpha\rangle$,
the expectation value \eqref{EV} reads
\be
\langle {\cal O}(t)\rangle={\rm Tr}(\rho_{DE} {\cal O })
+\sum_{\alpha\neq \beta}^{\cal D}
e^{i(E_\alpha-E_\beta)t}c_\alpha c^*_\beta \langle E_\alpha|{\cal O}|E_\beta\rangle\,,
\ee
where $\rho_{DE}$ is diagonal density matrix
\be
\rho_{DE}=
 \sum_{\alpha=1}^{\cal{D}} |c_\alpha|^2
 |E_\alpha\rangle\langle E_\alpha|\,.
\ee 
Here  ${\cal D}$ is the dimension of Hilbert space. Then, one can proceed to 
explore equilibrium and thermalization in this context which happens due to the possible phase cancellation at long times \cite{Serdnicki:1999} when  the expectation value of the operator may be given by the canonical ensemble 
${\rm Tr}(\rho_{DE} {\cal O })\approx{\rm Tr}(\rho_{th} {\cal O })$.

We note, however, that a Hamiltonian may be also put into a tridiagonal
form in which we could work in the Krylov basis. See {\it e,g}
\cite{{Dumitriu_2002},{Balasubramanian:2022dnj},{Balasubramanian:2023kwd}}. In this basis, we usually deal with  Lanczos
coefficients and thus we would expect that the properties of 
the quantum system can be also
described in terms of the spectrum of Lanczos coefficients. Indeed, the Lanczos 
spectrum has been used to study operator growth in many body systems
\cite{Parker:2018yvk} ( see also \cite{Barbon:2019wsy}). Recently,
it was also suggested in \cite{Balasubramanian:2023kwd} that the chaotic nature of a system may be 
described in terms of the Lanczos coefficients. More precisely, it was proposed 
that ``Quantum chaotic systems display a Lanczos spectrum well described by random matrix model.''


Here we would like to study quantum thermalization in the Krylov basis. In particular, we would like to understand
to what extent the nature of thermalization may be explored in this context (see also \cite{Bhattacharjee:2022qjw}). To proceed, let us first briefly review the recursive procedure producing 
the Krylov space for a given state in a quantum system ( see \cite{viswanath1994recursion} for review).

Starting with an initial state $|\psi_0\rangle$ 
in a  quantum system with a  time independent Hamiltonian $H$, the Krylov basis, $\{ |n\rangle, n=0,1,2,\cdots, 
{\cal D}_\psi-1\}$, can be constructed as follows. The first element of the basis is identified with the initial state $|0\rangle=|\psi_0\rangle$ (which we assume to be normalized)
and then  the other elements  are constructed, recursively,
as follows
\be\label{GS}
|\widehat{ n+1}\rangle=(H-a_n)|n\rangle -b_n|n-1\rangle\,,
\ee
where $|n\rangle =b_n^{-1}|{\hat n}\rangle$, and
\be\label{LC-state}
a_n=\langle n|H|n\rangle,\;\;\;\;\;\;\;\;\;b_n=\sqrt{\langle{\hat n}|{\hat n}\rangle}\,.
\ee
This recursive procedure stops whenever $b_n$ vanishes which occurs for 
$n={\cal D}_\psi\leq {\cal D}$ that is the dimension of Krylov space.
 Note that this procedure produces an orthonormal and ordered basis together with coefficients $a_n$ and $b_n$ known as Lanczos coefficients \cite{Lanczos:1950zz}.

Having constructed the Krylov basis, at any time the evolved state may be expanded in this basis 
\be
|\psi(t)\rangle=\sum_{n=0}^{{\cal D}_\psi-1}\phi_n(t)\,|n\rangle\,,\;\;\;\;{\rm with }\;\;\sum_{n=0}^{{\cal D}_\psi-1}|\phi_n(t)|^2=1\,,
\ee
where the wave function $\phi_n(t)$ satisfies the following Schr\"odinger equation
\be
-i\partial_t\phi_n(t)=a_n\phi_n(t)+b_{n}\phi_{n-1}(t)+b_{n+1}\phi_{n+1}(t)\,,
\ee
which should be solved with the initial  condition 
$\phi_n(0)=\delta_{n0}$.

Using the completeness of the energy eigenstates one may expand 
any element of the Krylov basis in terms of energy basis
\be
|n\rangle=\sum_{\alpha=1}^{{\cal D}} f_{n\alpha} |E_\alpha\rangle\,.
\ee
Note that since ${\cal D}_\psi\leq {\cal D}$, the expansion coefficient, $f_{n\alpha}$, is not 
necessary  inevitable and therefore, in general, energy eigenstates
cannot be expanded in terms of Krylov basis. Using the orthogonality condition 
of the Krylov basis one gets
\be
\sum_{\alpha=1}^{{\cal D}}f^*_{n\alpha}f_{m\alpha}=\delta_{nm}\,.
\ee
On the other hand from the equation \eqref{GS} one finds
\be\label{fna}
f_{n\alpha}E_\alpha=a_nf_{n\alpha}+b_{n+1}f_{n+1\alpha}+b_n
f_{n-1\alpha}\,,
\ee
that can be used to find $f_{n\alpha}$ in terms of $f_{0\alpha}=
c_\alpha$.

In this formalism  the expectation value of a local operator \eqref{EV} reads
\be\label{EVK}
\langle {\cal O}(t)\rangle=\sum_{n,m=0}^{{\cal D}_\psi-1}
\phi^*_n(t)\phi_m(t)\,O_{nm}
\ee
where $O_{nm}=\langle n|{\cal O}|m\rangle$ are  matrix elements of 
the operator ${\cal O}$ in the Krylov basis which, in general,
are complex numbers. Of course, the diagonal elements are real.

To find the  infinite time average of the 
corresponding operator one needs to
compute $C_{nm}$ given by 
\be
C_{nm}=\lim_{T\rightarrow \infty}\frac{1}{T}\int_0^Tdt\,\phi^*_n(t)\phi_m(t)\,,
\ee
that is essentially matrix elements of the diagonal density matrix
 in the Krylov basis $C_{nm}=\langle m|
 \rho_{DE}|n\rangle$, that is 
 \be
C_{nm}= \sum_{\alpha=1}^{\cal{D}} |c_\alpha|^2
 f_{m\alpha}f^*_{n\alpha}\,.
\ee

In this framework, inspired by ETH, we propose an ansatz for the matrix elements of typical operators in the Krylov basis to ensure that thermalization occurs within the system. To illustrate how the corresponding ansatz might be formulated, let us consider the ETH ansatz for the matrix elements of the operator ${\cal O}$ in energy eigenstates
 \cite{{Srednicki:1994mfb},{Serdnicki:1999}}\footnote{
 In this expression $S(\bar{E})$ is thermal entropy at energy $\bar{E}$ which is an extensive quantity and proportional to the size of the system. 
It is important to note that  $f$ and $f_O$ 
are smooth functions of their arguments.
 $R_{\alpha\beta}$ is a random real or complex variable with zero mean
 $\overline{R_{\alpha\beta}}=0$ and unit variance: $\overline{R^2_{\alpha\beta}}=1,\,
 \overline{|R_{\alpha\beta}|^2}=1$.},
\be\label{ETH}
\langle E_\alpha|{\cal O}|E_\beta\rangle = f(E_\alpha)
\delta_{\alpha\beta} + e^{-\frac{S(\bar{E})}{2}} f_{{\cal O}} (\bar{E},\omega) R_{\alpha\beta}\,,
\ee
where 
$\bar{E}=(E_\alpha+E_\beta)/2,\;\omega=E_\alpha-E_\beta$.
This expression can be utilized to propose an ansatz for the matrix elements of the operator ${\cal O}$ in the Krylov basis through the following relation
 \be\label{Kb-Eb}
\langle n|{\cal O}|m\rangle=\sum_{\alpha,\beta
=1}^{{\cal D}}f^*_{n\alpha}f_{m\beta}\;\;\langle E_\alpha|{\cal O}|E_\beta\rangle\,.
\ee
To proceed, one can promote the function $f$ to an operator by replacing the energy with the Hamiltonian
\be
f(E_\alpha)\rightarrow \hat{f}(H)\,.
\ee
This allows 
to  express the ETH ansatz  as follows
\be
\langle E_\alpha|{\cal O}|E_\beta\rangle =\langle E_\alpha| \hat{f}(H)|E_\beta\rangle+{\cal O}(e^{-S/2})\,.
\ee
Here we have utilized the fact that  
$\hat{f}(H)|E_\beta\rangle=f(E_\beta)|E_\beta\rangle$. 
Plugging this expression into equation \eqref{Kb-Eb}
one gets
\be
\langle n|{\cal O}|m\rangle =\langle n| \hat{f}(H)|m\rangle+{\rm suppressed\; terms}\,.
\ee
In this equation, the suppressed terms correspond to the exponentially suppressed terms in the original ETH ansatz. 
Using the fact that
$\langle n|m\rangle=\delta_{nm}$ and
\be\label{bb}
\langle n|H|m\rangle=a_n\delta_{n\;m}+b_{m+1}\delta_{n\; m+1}+
b_m \delta_{n\; m-1},
\ee
it becomes clear that the matrix representation of the operator 
${\cal O}$ in the Krylov basis is not diagonal. However, we note that while off-diagonal elements also appear in leading order in this expression, we generally would not expect significant contributions from all off-diagonal matrix elements.

To understand this, we recognize that for thermalization
to occur beside the ETH ansatz, one must further assume that
$f$ is a smooth and slowly varying function of $E_\alpha$. Additionally, the initial state must be sufficiently localized within a narrow energy window—specifically, the variance of energy should be much smaller than the energy expectation value of the initial state. This  justifies the neglecting of higher-order terms in the following Taylor expansion\cite{Serdnicki:1999}\footnote{More precisely, one 
assumes that $(\Delta E)^2 f''(E)/f(E)\ll 1$
with $\Delta E$ being the variance of energy\cite{Serdnicki:1999}.}
\be
f(E_\alpha)\approx f(E)+
(E_\alpha-E)\,f'(E),
\ee
where $E=\langle \psi_0|H|\psi_0\rangle$ and, $f(E)$
is the expectation value predicted by the (micro)canonical ensemble
${\rm Tr}(\rho_{th}{\cal O})\approx f(E)$. Here  ``prime'' denotes 
derivative with respect to $E_\alpha$. In this approximation, the ETH ansatz \eqref{ETH} reads
\be
{\cal O}_{\alpha\beta}\approx f(E)
\delta_{\alpha\beta} + f'(E)\,\langle E_\alpha|H-E|E_\beta\rangle
+{\cal O}(e^{-\frac{S}{2}})\,,
\ee
resulting in the following expression for the matrix elements 
in the Krylov basis
\bea\label{KTH10}
{\cal O}_{nm}&\approx& f(a_0)\delta_{nm} 
+ f'(a_0)\,\langle n|H-a_0|m\rangle
\cr &&+{\rm suppressed\; terms}\,,
\eea
which shows that the matrix representation of a typical
observable in the Krylov basis is essentially tridiagonal
in which the off-diagonal elements adjacent to the diagonal, denoted as ${\cal O}_{nn+1}$, are proportional to the Lanczos coefficients $b_n$ (see equation \eqref{bb}).  Note that the suppressed terms, which contain $b_n$-dependent factors, are of order ${\cal O}(\Delta E^2)$.

Regarding "suppressed terms," one can follow a similar approach as with the ETH ansatz to estimate their order of magnitude \cite{DAlessio:2015qtq}. To proceed,
following \cite{Mori:2017qhg},
we note that  using the completeness of the Krylov basis, the
inequality the fact $\langle n|{\cal O}^2|n\rangle\leq |{\cal O}|^2$ maybe
written as $\sum_m |{\cal O}_{nm}|^2\leq |{\cal O}|^2$. Moreover, since 
$|{\cal O}_{nn}|^2$ is always positive one may write
\be\label{off-O}
\sum_{m (\neq n)=0}^{{\cal D}_\psi-1} |{\cal O}_{nm}|^2\leq 
|{\cal O}|^2\,.
\ee
where $|{\cal O}|$ denotes the operator norm\footnote{
The norm is defined by $|{\cal O}|={\rm sup}_\psi\sqrt{\langle \psi|
{\cal O}^\dagger {\cal O}|\psi\rangle}$ \cite{Mori:2017qhg}.
}, and ${\cal D}_\psi$ is the dimension of the Krylov space associated with the initial state $\psi_0$.
Moreover, if we assume that the off-diagonal matrix elements ${\cal O}_{nm}$ are smooth and vary slowly, they can be considered to be almost constant so that one finds
\be
\sum_{m (\neq n)=0}^{{\cal D}_\psi-1} |O_{nm}|^2\approx |O_{nm}|^2\sum_{m (\neq n)=0}^{{\cal D}_\psi-1}=|O_{nm}|^2 {\cal D}_\psi\,.
\ee
By making use of this approximation one can derive an expression that captures the behavior of off-diagonal elements  to the leading order. Indeed from 
 equation \eqref{off-O}, we one gets
\be
|{\cal O}_{nm}|\leq \frac{|{\cal O}|}{\sqrt{{\cal D}_\psi}}\,.
\ee
 It is essential to highlight that the aforementioned condition pertains to the contributions of "suppressed terms" to off-diagonal matrix elements. Additionally, there is a contribution to these off-diagonal matrix elements from leading terms, as indicated in equation \eqref{KTH10}.

Inspired by the above observations one can propose an ansatz for the matrix elements of typical operators in the Krylov basis as follows 
\be\label{KTH1}
\langle n|{\cal O}|m\rangle =\langle n| \hat{f}(H)|m\rangle+\frac{1}{\sqrt{{\cal D}_\psi}}  f_{{\cal O}} (a_{nm}) R_{nm}\,,
\ee
where $a_{nm}=\langle n|H|m\rangle$.
 Here $R_{nm}$ is a random real or complex variable with zero mean
 $\overline{R_{nm}}=0$ and unit variance: $\overline{R^2_{nm}}=1,\,
 \overline{|R_{nm}|^2}=1$. 

The equation \eqref{KTH1} imposes a condition on the matrix elements of 
the operator ${\cal O}$ in the Krylov basis, which can be seen as an 
ansatz for these matrix elements to ensure thermalization. This is 
referred to as the KTH ansatz. It is worth noting that, in practice, for 
typical chaotic systems, the leading-order terms of the KTH ansatz are 
actually represented by those in equation \eqref{KTH10}.
In fact by substituting this expression into equation \eqref{EVK} one finds 
\bea
&&\langle {\cal O}(t)\rangle=\sum_{n,m=0}^{{\cal D}_\psi-1}
\phi^*_n(t)\phi_m(t)\,\langle n| \hat{f}(H)|m\rangle\\
&&\;\;\;\;\;\;\;\;\;\;\;\;\;+\frac{1}{\sqrt{{\cal D}_\psi}}\sum_{n,m=0}^{{\cal D}_\psi-1}
\phi^*_n(t)\phi_m(t)  f_{{\cal O}} (a_{nm}) R_{nm}\,.\nonumber
\eea
Note that this equation indicates that the off-diagonal terms, which contain $b_n$-dependent terms, remain exponentially suppressed, scaling as ${\cal O}(e^{-S/2})$.

One can use \eqref{KTH10} to simplify the first line of the
above equation. It is clear that the contribution of off-diagonal term in \eqref{KTH10} vanishes\footnote{Note that $\sum_{n,m=0}^{{\cal D}_\psi-1}
\phi^*_n(t)\phi_m(t)\,\langle n| H|m\rangle=E$ and by definition $a_0=E$.}, whereas from the first term and taking into account that $\sum_n|\phi(t)|^2=1$ 
one arrives at 
\be
\langle {\cal O}(t)\rangle\approx f(a_0)+
 \frac{1}{\sqrt{{\cal D}_\psi}}  
 \sum_{n,m=0}^{{\cal D}_\psi-1}
\phi^*_n(t)\phi_m(t)\,f_{{\cal O}} (a_{nm})
 R_{nm},
\ee
which results in 
\be\label{Otime}
\langle {\cal O}(t)\rangle\approx {\rm Tr}(\rho_{th}{\cal O})+
{\rm small\; fluctuations},
\ee
as expected.  Here we have used that the system thermalizes so that the expectation value of the operator is equal to thermal expectation value $f(a_0)={\rm Tr}(\rho_{th}{\cal O})$.
Therefore, even though the matrix is tridiagonal, the main contribution is primarily determined by the diagonal elements, which represent the expectation value of the corresponding operator as derived from the (micro)canonical ensemble.
It is also worth noting that in our context, all physical results are fundamentally determined by two parameters: $a_0$ and $b_1$, both of which carry physical significance. Specifically, 
$a_0 = E$ represents the energy of the initial state, while $b_1 = \Delta E$ denotes its variance.. 

To examine the KTH behavior of local operators (observables) we will compute 
matrix elements of the  operator $S_x=\sum_{i=1}^N\sigma^x_i$, which is  magnetization in the 
$x$ direction, for the
Ising model \eqref{Ising} with $h=0.5,\,g=-1.05$ where 
the model is non-integrable.

\begin{figure}[h!]
	\begin{center}
		\includegraphics[width=0.49\linewidth]{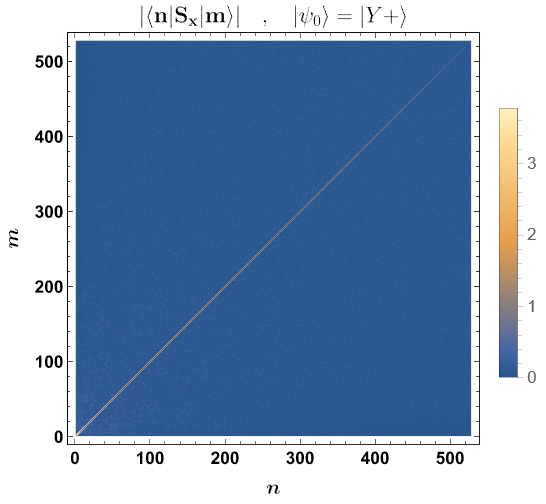}
\includegraphics[width=0.49\linewidth]{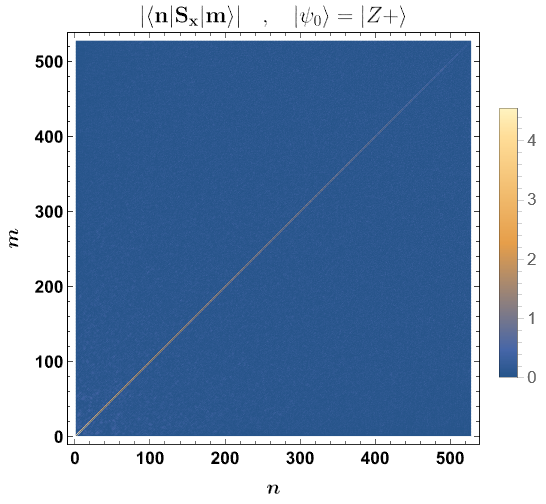}
  \end{center}
	\caption{ Matrix elements of the operator  $S_x=\sum_i\sigma_i^x$ in Krylov basis for 
 cases where the initial state is $|Y+\rangle$ (left) and 
 $|Z+\rangle$ (right). Darker  points represent  the matrix elements that are closed
 to zero. Although, due to its resolution, it might not be clear
 for these plots,
 these matrices are not diagonal and, indeed, they are tridiagonal
(see figure \ref{fig:expectation-value-X}).}
	\label{fig:expectation-value}
\end{figure}

The corresponding matrix elements for two different initial states
$|Y+\rangle$ and $|Z+\rangle$ are presented in figure 
\ref{fig:expectation-value} for $N=10$.  
From this figure, one observes that the matrix elements $O_{nm}$ exhibit the desired behavior as suggested by KTH. To highlight the behavior of matrix elements we have presented the absolute value of them, so that in figure \ref{fig:expectation-value} the dark 
points correspond to vanishing elements.

\begin{figure}[h!]
	\begin{center}
		\includegraphics[width=0.49\linewidth]{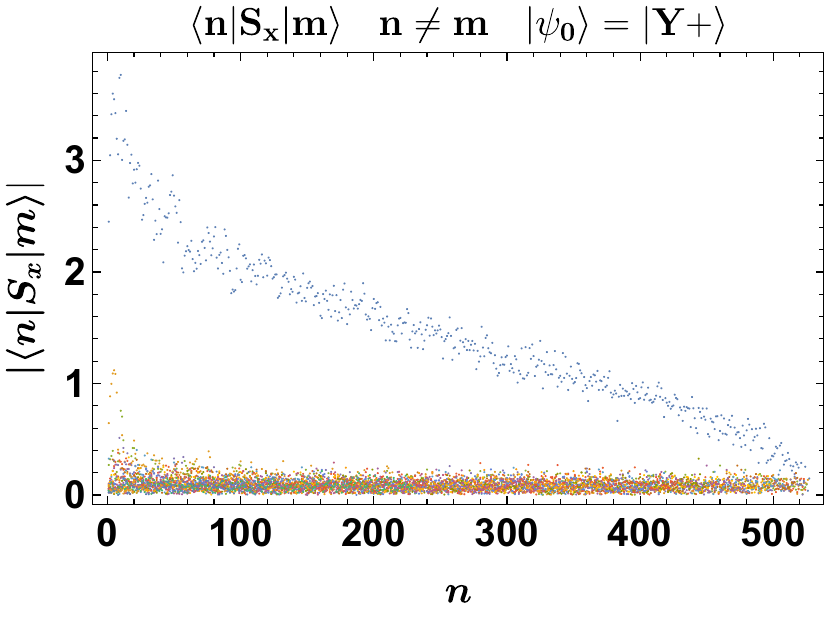}
\includegraphics[width=0.49\linewidth]{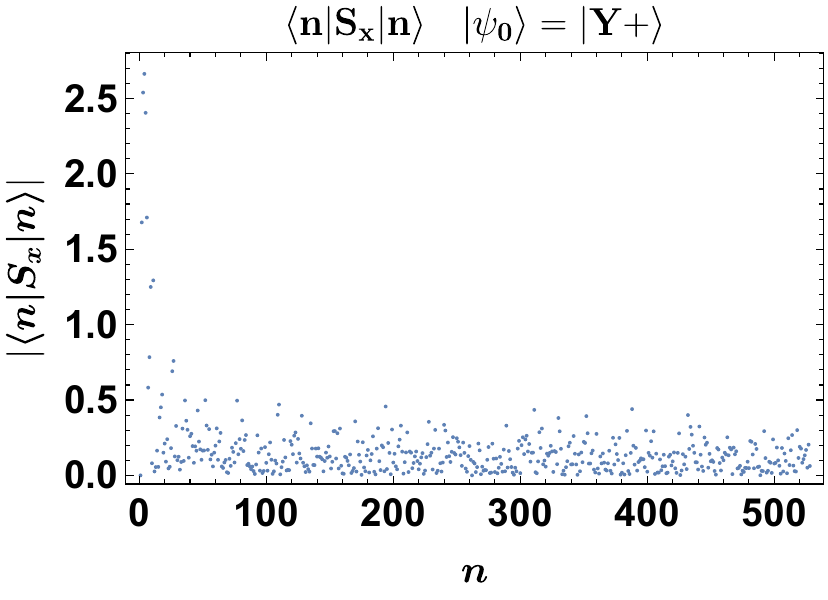}
\includegraphics[width=0.49\linewidth]{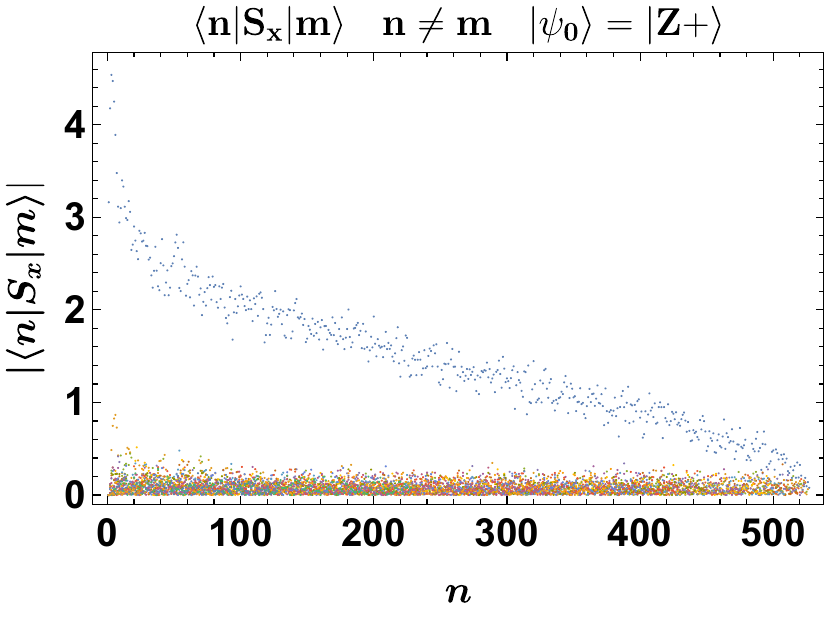}
\includegraphics[width=0.49\linewidth]{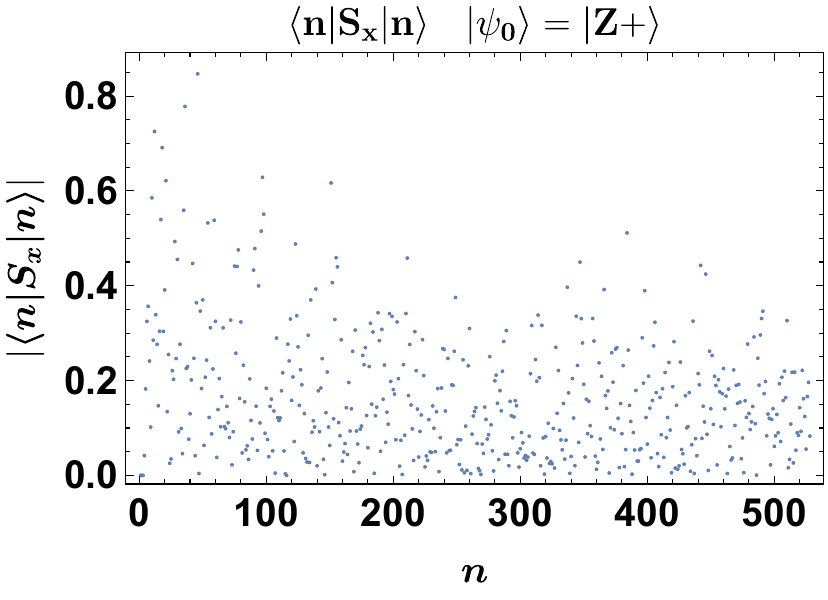}
  \end{center}
	\caption{Actual (absolute)  values of matrix elements of the operator  $S_x=\sum_i\sigma_i^x$ in Krylov basis for 
 cases where the initial state is $|Y+\rangle$ (up) and 
 $|Z+\rangle$ (down).  Blue points in left panels denote 
 $(S_x)_{n,n+1}$ elements.
}
	\label{fig:expectation-value-X}
\end{figure}

 To further explore KTH we have also presented the actual values 
 of matrix elements of $S_x$ in the Krylov basis for different initial states in figure \ref{fig:expectation-value-X} where one can observe
 that $(S_x)_{n,n+1}$ is significantly greater than other elements. Note also that, diagonal elements ${\cal O}_{nn}$ are not entirely given by $f(a_0)$ and, in fact,
 they appear in a certain combination of $f(a_0)+(a_n-a_0)
f'(a_0)$ which might be small,
even though the expectation value, itself, could be relatively large.

\begin{figure}[h!]
	\begin{center}
		\includegraphics[width=0.49\linewidth]{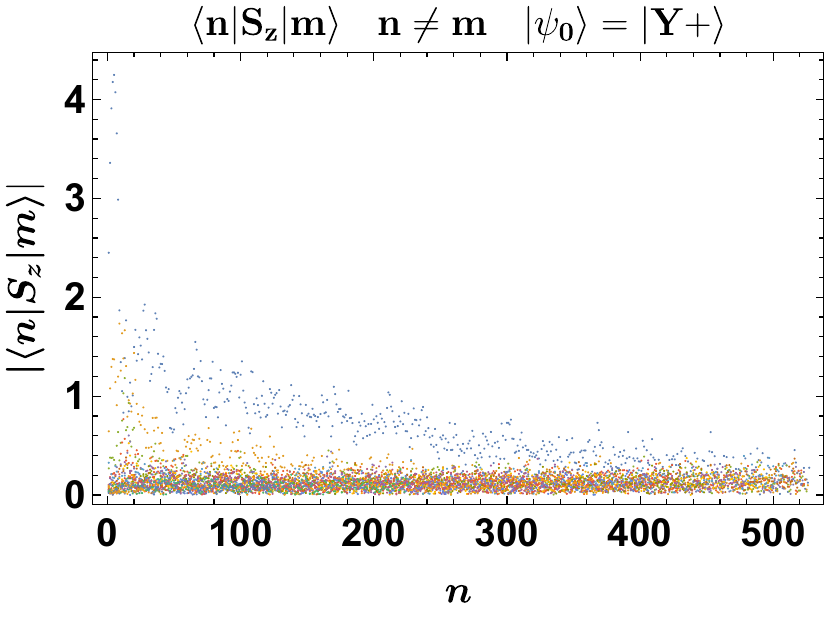}
\includegraphics[width=0.49\linewidth]{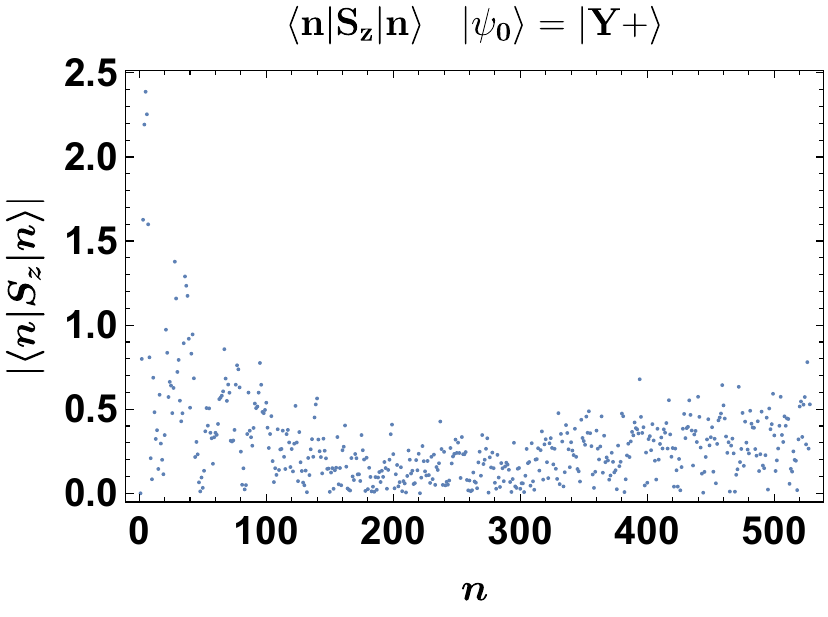}
\includegraphics[width=0.49\linewidth]{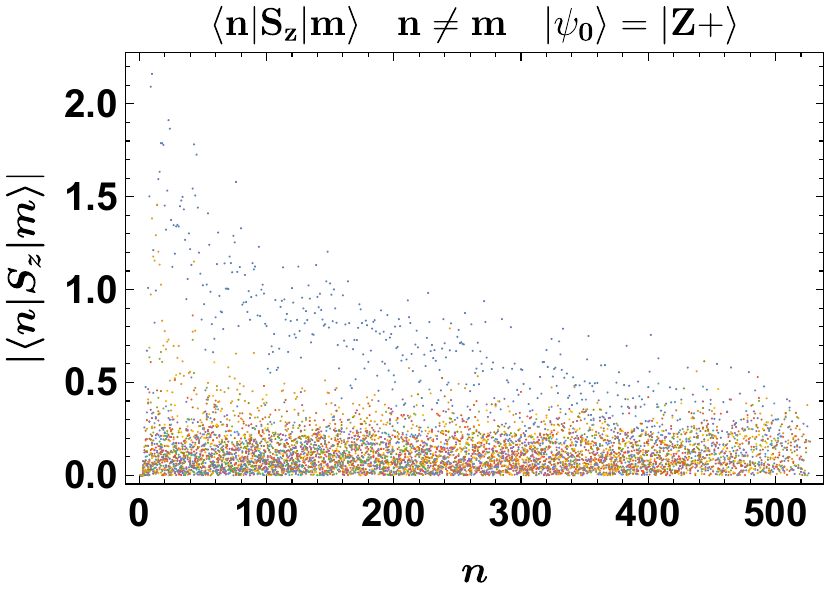}
\includegraphics[width=0.49\linewidth]{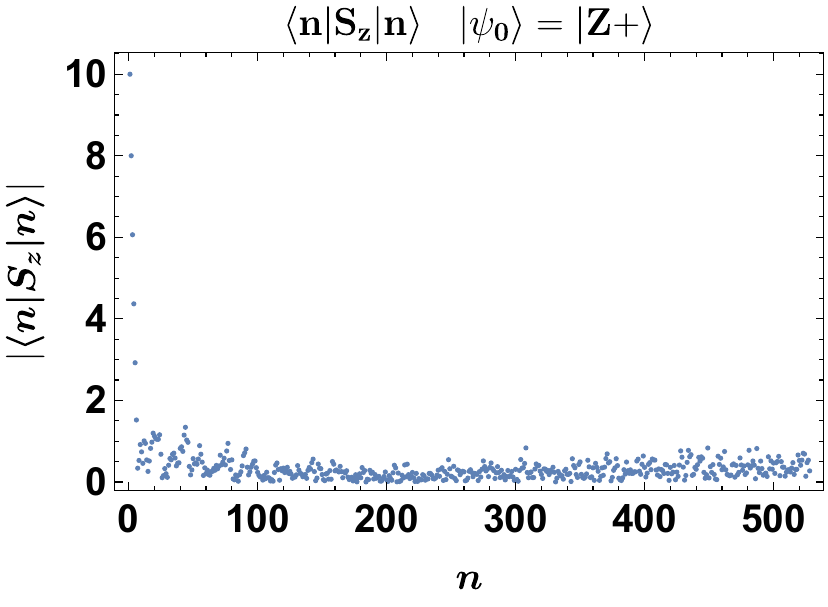}
  \end{center}
	\caption{Actual (absolute)  values of matrix elements of the operator  $S_z=\sum_i\sigma_i^z$ in Krylov basis for 
 cases where the initial state is $|Y+\rangle$ (up) and 
 $|Z+\rangle$ (down).  Blue points in the left panels denote 
 $(S_z)_{n,n+1}$ elements.
}
	\label{fig:expectation-value-Z}
\end{figure}

 We have also computed matrix elements of the magnetization 
 in the $z$ direction, $S_z=\sum_{i=1}^N\sigma^z_i$, for several initial states 
 specified  by different  $\theta$ and $\phi$ and we have
 found  the same pattern as that in 
 figure \ref{fig:expectation-value} and \ref{fig:expectation-value-X}.
 In particular, numerical results for initial states $|Y+\rangle$
 and $|Z+\rangle$ are depicted in figure\ref{fig:expectation-value-Z}.

\begin{figure}[h!]
	\begin{center}
		\includegraphics[width=0.49\linewidth]{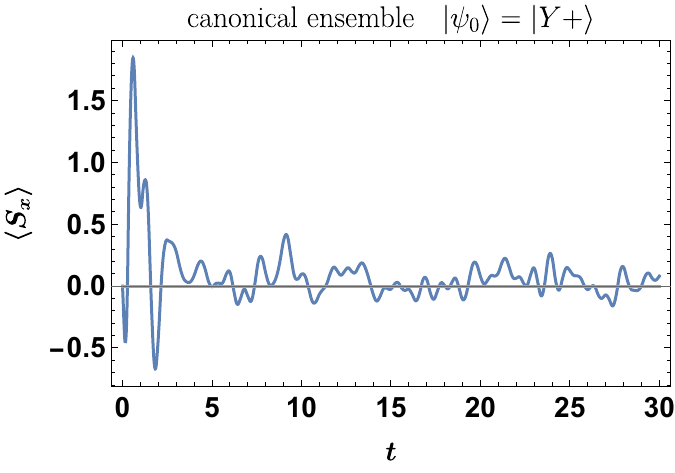}
\includegraphics[width=0.49\linewidth]{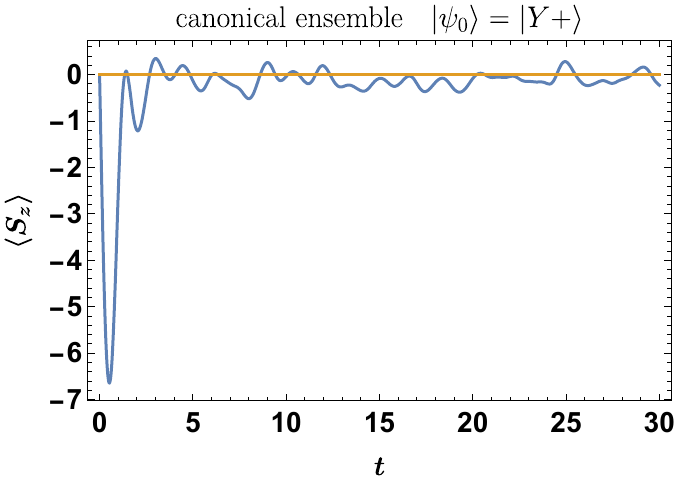}
\includegraphics[width=0.49\linewidth]{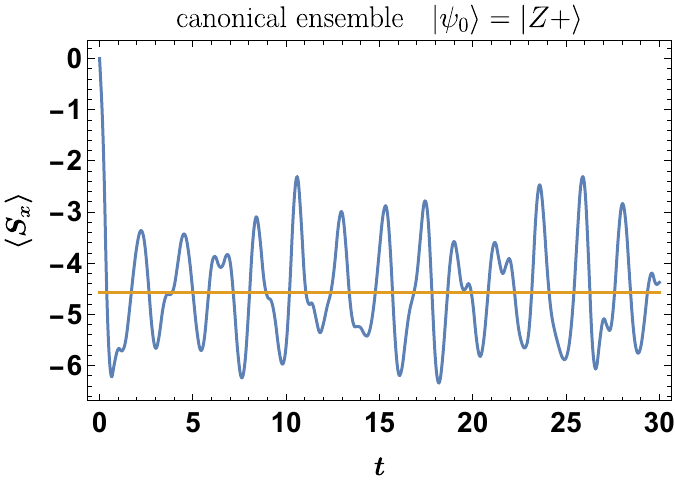}
\includegraphics[width=0.49\linewidth]{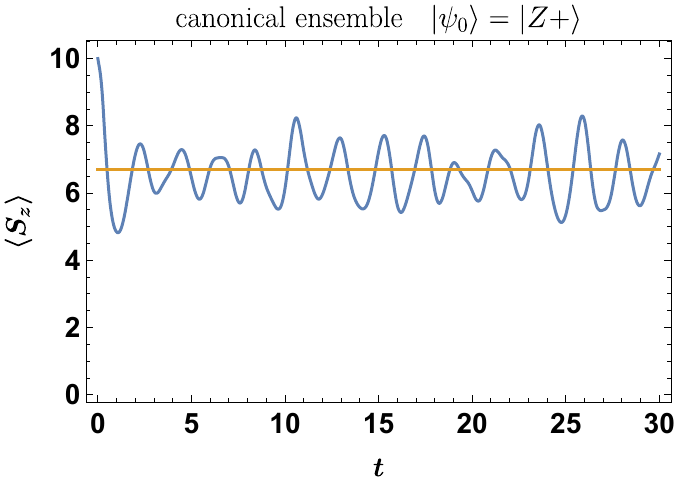}
  \end{center}
	\caption{Time evolution of the expectation value of $S_x$ and $S_z$ for two different initial states $|Y+\rangle$ (up) 
 and $|Z+\rangle$ (down). The straight lines
 represent the value predicted by the canonical ensemble. 
}
	\label{fig:expectation-value-S(t)}
\end{figure}

  It is also illustrative to explicitly  compute time 
  evolution of the operator we considered above to see 
  how they actually follow the general behavior given by 
  the equation \eqref{Otime}. The results are depicted 
  in figure \ref{fig:expectation-value-S(t)}. The straight brown lines
  in these plots represent the long time expectation value
  predicted by the canonical ensemble which is equal to
  the infinite time average of the
  corresponding expectation value: ${\rm
  Tr}(\rho_{DE}S_{x,z})\approx{\rm
  Tr}(\rho_{th}S_{x,z})$, that is necessary for thermalization to occur. 
Although thermalization occurs for both initial states, from  this figure one observes that the nature of thermalization should be different for these
states. While $|Y+\rangle$
exhibit string thermalization, for $|Z+\rangle$ it is weak.

 To further explore the KTH behavior we have also done the same 
 computations for the cases where $gh=0$ in which the model \eqref{Ising} is integrable.
 An immediate observation we have made is that in integrable cases the dimension of Krylov 
 space reduces significantly\footnote{Similar
 observation has been already  made in the 
 context of  operator Krylov complexity in 
\cite{Rabinovici:2020ryf} (see also
 \cite{Caputa:2024vrn}).} which may also depend on the initial state. For example, for $N = 10$, where  the dimension of the full Hilbert space is $2^{10}$, in the chaotic case, the dimension of the Krylov space is 529\footnote{It is known that the Hamiltonian \eqref{Ising} has a parity symmetry which is essentially reflection symmetry about the center of the chain. It is straightforward to see that 
the initial state \eqref{initial} has positive parity which in turn indicates that the obtained Krylov subspace should be a subspace with positive parity. Actually, the positive parity subspace has 528 dimensions, as expected, which is equal to the dimension of Krylov space.}. In the integrable case with parameters $h = 0$  and $g = -1.05$, the
corresponding dimensions are 463 for the initial states $|Y+\rangle$ 
 and $|Z+\rangle$, and 253 for $|X+\rangle$.

\begin{figure}[h!]
	\begin{center}
		\includegraphics[width=0.55\linewidth]{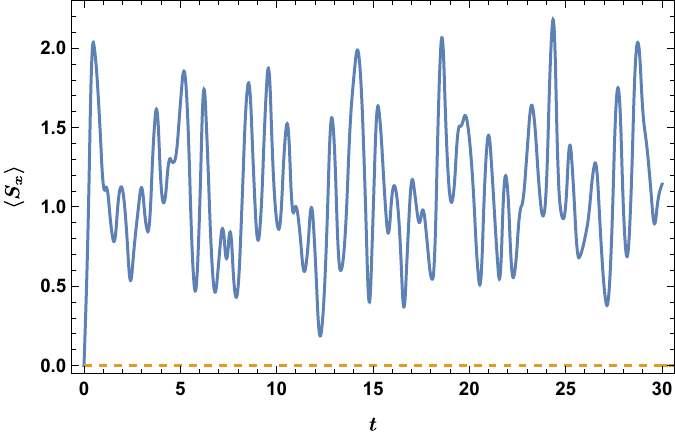}
  \end{center}
	\caption{Time evolution of the expectation value of $S_x$ for initial state $|Y+\rangle$ for an integrable case 
 where $h=0$. The brown dashed line represents a value predicted by 
 the canonical ensemble whose energy is the same as that of the initial state $|Y+\rangle$.
  }
	\label{fig:expectation-value-S-Integ}
\end{figure}

In order to have a better statistic we have considered the case 
where $h=0,\,g=-1.05$. For this case, we have computed matrix 
elements of different operators in the Krylov basis 
 for different initial states. We have found that,  although 
 for some special cases, the corresponding matrix elements have 
 almost similar patterns as that in figure \ref{fig:expectation-value},  it is not a generic behavior and typically they exhibit 
 non-universal behavior. More importantly, in this case 
 the long time average cannot be approximated by a canonical 
 ensemble (see, for example, figure \ref{fig:expectation-value-S-Integ}).

\section{Krylov space and nature of thermalization}

As we have already maintained  although both initial states 
$|Y+\rangle$ and $|Z+\rangle$  exhibit almost the same pattern for the operator matrix elements
(see figures \ref{fig:expectation-value-X}
and \ref{fig:expectation-value-Z}) indicating that thermalization 
occurs in both states, it is evident for  figure \ref{fig:expectation-value-S(t)} that the nature of thermalization for these two states
must be different, as we discussed in the previous 
section. In this section, we would like to study how the nature of 
thermalization, being weak or strong, can be probed in the context
of Krylov space. 

From equation \eqref{Otime} one finds that the nature of 
thermalization should be controlled by the second term which 
is essentially given by a summation over $\phi_n(t)$'s. 
On the other hand, $\phi_n(t)$ can be evaluated, recursively, 
from $\phi_0(t)$ using  Lanczos coefficients. Therefore, the nature of thermalization should be reflected in these quantities. Based on this insight, we will propose various quantities within the framework of Krylov space that could serve as indicators for the nature of thermalization.


\subsection{Variance of Lanczos coefficients}

From the Krylov basis construction, it is evident that the Lanczos coefficients should encapsulate information about both the model's dynamics and initial state, making them a suitable candidate for probing the nature of thermalization.

\begin{figure}[h!]
		\includegraphics[width=0.48\linewidth]{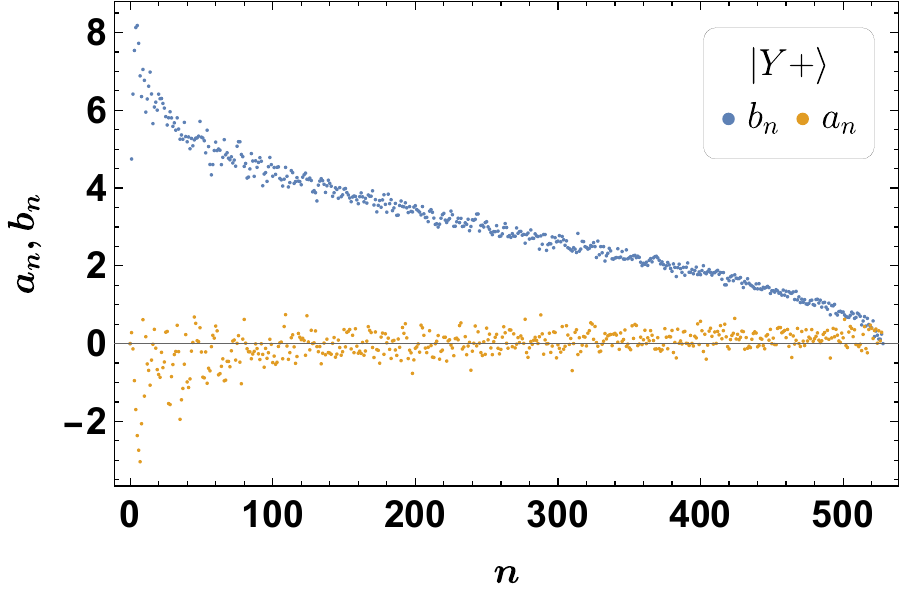}
		\includegraphics[width=0.48\linewidth]{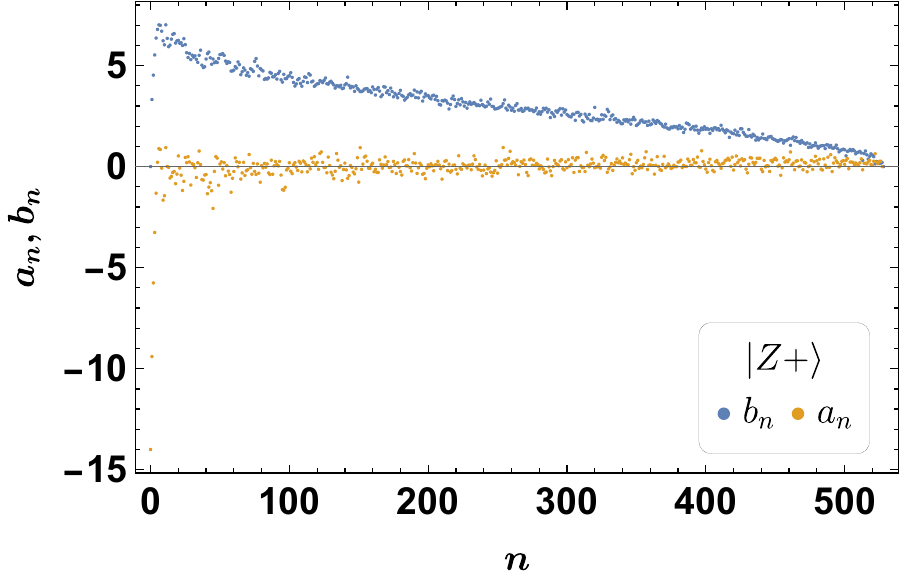}
  \includegraphics[width=0.48\linewidth]{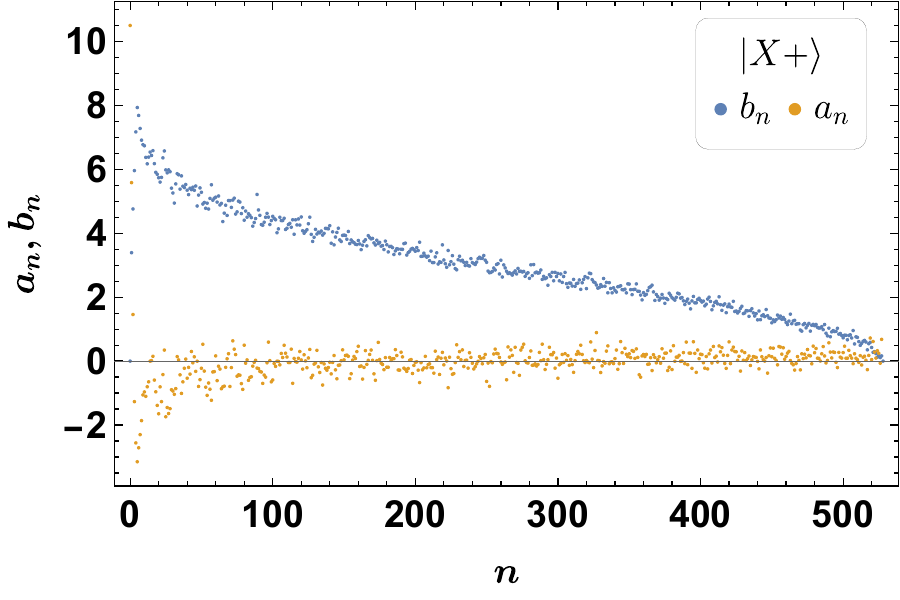}
	\caption{ Lanczos coefficients $a_n, b_n$ of three initial
  states $|Y+\rangle, |Z+\rangle,
 |X+\rangle$.  $b_n$ and $a_n$
 are shown with blue and brown circles, respectively. The numerical results are 
 presented for $N=10$ in which the dimension of Krylov space for a generic
 initial state is about $528$.
}
	\label{fig:LC}
\end{figure}

To explore this idea, let us begin by calculating the Lanczos coefficients for three different initial states that we have discussed in the previous section.\footnote{Lanczos coefficients for the model 
under consideration have also been computed in
\cite{Noh_2021, Trigueros:2021rwj, Espanol:2022cqr,Bhattacharya:2023zqt,Bhattacharya:2022gbz, Scialchi:2023bmw}.}. The results for $N=10$
are shown in  figure \ref{fig:LC}

Although one could recognize some differences 
among these three 
plots, the differences are not substantial. Essentially,
the Lanczos coefficients exhibit qualitatively similar patterns across all three cases. As a result, one might conclude that the straightforward behavior of Lanczos coefficients may not offer a distinct metric to differentiate between these three cases.

We note, however, that 
a better quantity which might be more sensitive to the initial state 
is the variance of  Lanczos coefficients. Indeed, the variance of 
Lanczos coefficients
has been considered in \cite{Scialchi:2023bmw} as a measure to 
probe whether a system is chaotic or integrable. 

Let us recall that for a collection of $M$ numbers, $s_i$, the variance may be defined as 
follows
\be
{\rm Var}(s_i)=\frac{1}{M}\sum_{i=1}^M(s_i-\bar{s})^2\,
\ee
where  $\bar{s}$ is the mean value. In what follows we would like
to compute the variance of Lanczos coefficients $a_n$ and $b_n$.
Actually, if one computes the variance of Lanczos coefficients for three
initial states considered before, one observes that they differ significantly.

\begin{figure}[h!]
	\begin{center}
		\includegraphics[width=0.47\linewidth]{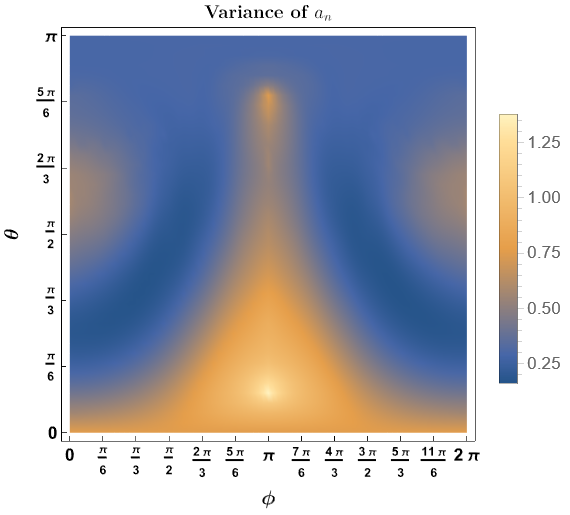}
\includegraphics[width=0.47\linewidth]{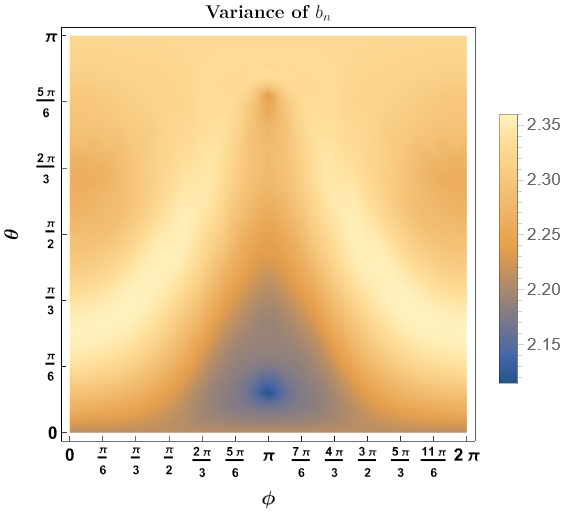}
  \end{center}
	\caption{Variance of Lanczos coefficients $a_n$ (left) and $b_n$ (right).
The numerical results are done for $N=10$ spins.  
}
	\label{fig:var}
\end{figure}

 More generally, one could compute the variance of Lanczos coefficients 
 associated with the general initial state given by \eqref{initial}.  In figure \ref{fig:var} 
 we have presented the numerical results  for  the variance of
 $a_n$ and $b_n$ as a  function of $\theta$ and $\phi$ for $N=10$. 

Clearly, there is an obvious correlation  between behaviors of the effective inverse temperature, the absolute value of the density of energy
(or normalized energy)  and variance of 
Lanczos coefficients (see figure \ref{fig:Ising1}). Generally, one observes that for regions where 
the effective inverse temperature is
small the variance of $a_n$ ($b_n$) is also small (large). The 
variance of
$a_n$ ($b_n$) becomes larger (smaller) as we move
away  from $\beta=0$ 
regions. 
Another observation we have made is that being positive, the variance is not 
sensitive to  the sign of $\beta$ and only the absolute value of 
it matters.

We note, however, that the behavior of variances is not exactly 
the same as that of effective inverse temperature.  
Indeed, even though one can recognize the lower part of the
ring of zero 
$\beta$, the upper part is not apparent in the plots of variances, though 
there is a trace of the ring.
More precisely, although from the behavior of $\beta$
or density of energy one would expect to see states with strong thermalization 
are localized near the ring of zero $\beta$, 
the behavior of the variance suggests that strong thermalization
for states with $\theta \gtrsim \frac{2\pi}{3}$
does not necessarily located near the ring of zero $\beta$ and 
rather they almost uniformly distribute around $\theta\approx \pi$.

Notably, along the symmetric axis at \( \phi = \pi \), while the effective inverse temperature and the absolute value of the energy density decrease almost monotonically from \( \theta = \frac{\pi}{6} \) to \( \theta = \pi \), the variance of \( a_n \) exhibits a minimum around \( \theta \approx \frac{5\pi}{6} \). This suggests that the state \( |\frac{5\pi}{6},\pi\rangle \) is among those with the weakest thermalization.

\begin{figure}[h!]
	\begin{center}
		\includegraphics[width=0.55\linewidth]{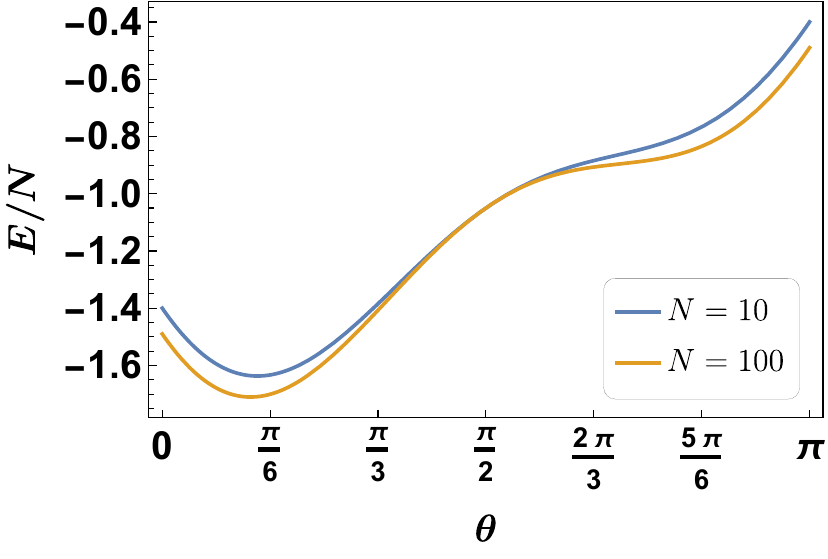}
		  \end{center}
	\caption{Density of energy (left) and its derivative (right)
 for  $\phi=\pi$ slice for different $N=10,100$. Indeed, they show
 nontrivial behavior around $\theta=\frac{5\pi}{6}$, though it is not as pronounced as that in the variance of $a_n$.
}
	\label{fig:E-phi=pi}
\end{figure}

To validate our numerical results, we can leverage the exact analytic expression for the energy. This allows us to compute the energy density for the $\phi=\pi$ slice for arbitrary $N$. In figure 
\ref{fig:E-phi=pi}, we present the energy density for $N=10$ and 
$N=100$ to further examine its behavior.

From this figure, we observe that the energy density exhibits non-trivial behavior around $\theta = \frac{5\pi}{6}$, though this is less pronounced than the variance of $a_n$. Using the explicit form of the energy density for large $N$ given in equation \eqref{ED}, we can compute its derivative with respect to $\theta$ for the 
$\phi = \pi$ slice
\be
E' = \frac{N}{2} \left(\sin\theta + 4\cos\theta \sin\theta - \frac{21}{10}\cos\theta\right)\,,
\ee
where the prime indicates differentiation with respect to $\theta$. The potential minima can be identified by solving the equation
\be
\sin\theta + 4\cos\theta \sin\theta - \frac{21}{10}\cos\theta = 0\,.
\ee
It is straightforward to see that this equation has only one solution for $\theta \in [0, \pi]$. Therefore, while the second point may appear significant, there is no true minimum around $\theta = \frac{5\pi}{6}$. Remarkably, for large $N$, the position of the minimum remains independent of $N$.

In conclusion, we find a correlation between the behavior of the effective inverse temperature and the variance of the Lanczos coefficients. However, notable discrepancies persist between the two, suggesting that the variance of the Lanczos coefficients may not be a reliable indicator of the nature of thermalization, despite its potential to capture certain aspects of it.

\subsection{ Infinite time average of Krylov complexity }

Working with Krylov basis we note that there is rather a special operator in Krylov space  whose matrix elements are proportional to Kronecker delta. More precisely, consider the number operator defined by
 \be
 {\cal N}=\sum_{n=0}^{{\cal D}_\psi-1} \,n|n\rangle\langle n|,
 \ee
that is obviously diagonal in Krylov basis, ${\cal N}_{nm}=n\delta_{nm}$. We note that the expectation value of the number operator, actually, computes Krylov complexity \cite{Parker:2018yvk}\footnote{See also \cite{{Avdoshkin:2019trj},{Dymarsky:2019elm},{Dymarsky:2021bjq},{Kim:2021okd},{Adhikari:2022whf},{Camargo:2023eev},{Huh:2023jxt}} for related works.} 
\be\label{KC}
{\cal C}=\langle {\cal N}(t)\rangle =\sum_{n=0}^{{\cal D}_\psi-1}
\;n\;|\phi_n(t)|^2\,,
\ee
that saturates  at very late times where the Lanczos coefficients vanish \cite{Rabinovici:2022beu}. The Krylov complexity is an interesting quantity which 
relies on both the initial state and the Hamiltonian, akin to Lanczos 
coefficients. 

The evolution of maximally entangled states in the Krylov basis has been studied in \cite{Erdmenger:2023wjg}, revealing that the growth and subsequent saturation of Krylov complexity are common features of many-body systems, regardless of their chaotic or integrable nature. As a general consequence, we would expect to observe the following behavior at late times
\be
{\cal C}=\overline{{\cal C}}
+{\rm small\; fluctuations}\,,
\ee
where $\overline{{\cal C}}$ is the infinite time average of 
the Krylov complexity given by 
\be
\overline{{\cal C}}=\lim_{T\rightarrow\infty}\frac{1}{T}\int_0^T
\langle {\cal N}(t)\rangle\, dt={\rm Tr}(\rho_{DE}{\cal N})\,,
\ee

It was conjectured that one could probe the system's dynamics by examining the infinite time average of Krylov complexity, with chaotic models exhibiting higher values \cite{Rabinovici:2021qqt}. 
In fact,  in an explicit Ising model, it has been demonstrated  in  \cite{Rabinovici:2021qqt}
that the 
infinite time average of the Krylov complexity increases as 
one transitions from an integrable model to a chaotic 
one. 
However, it is important to note that 
in addition to the dynamics the infinite time average of Krylov complexity also depends on the initial states.
In fact, the infinite time average of Krylov complexity for chaotic systems may or may not exceed that of integrable models 
\cite{Scialchi:2023bmw, Trigueros:2021rwj, Espanol:2022cqr}.

\begin{figure}[h!]
	\begin{center}
		\includegraphics[width=0.8\linewidth]{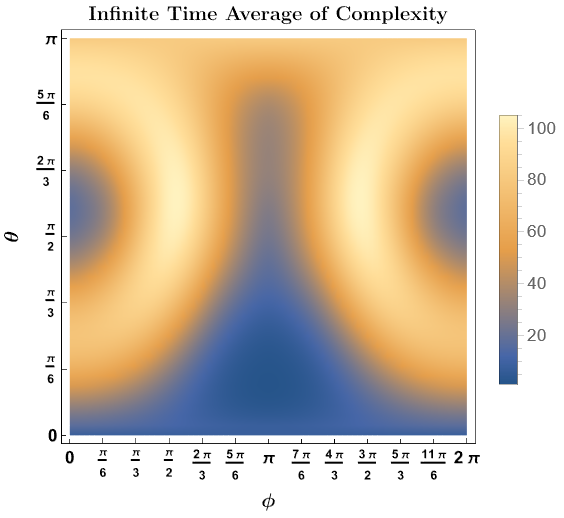}  
		  \end{center}
	\caption{Infinite time average of complexity for states
 associated with the initial state \ref{initial}. The numerical 
 computation is done for $N=9$.
}
	\label{fig:complexity}
\end{figure}

We have leveraged this insight to propose the infinite time average of Krylov complexity as a measure to probe the nature of thermalization.
It is straightforward to compute the infinite time average of 
complexity for states associated with initial states
\eqref{initial}. The numerical result for $N=9$ is depicted in figure \ref{fig:complexity}.

Interestingly, the resulting pattern aligns perfectly with the absolute value of the energy density and the effective inverse temperature. Specifically, we have observed that states exhibiting strong thermalization have a higher complexity saturation value than those with weak thermalization. In conclusion, for the model described in equation \eqref{Ising}, we find that as thermalization intensifies, the complexity saturation value also increases. We hypothesize that similar conclusions may apply to generic models.


\subsection{Inverse participation ratio }

It is worth mentioning that by making use of the inverse participation 
 ratio \cite{Short:2011pvc} the nature of weak or strong 
 thermalization of certain XY Ising model has been studied in\cite{Prazeres:2023hce}. Thus it is worth looking at this quantity for our model too.
 
 Consider a state whose expansion in the energy eigenstates is 
 $ |\psi\rangle=\sum_{\alpha=1}^{\cal D} c_\alpha |E_\alpha\rangle$, 
 where $c_\alpha=\langle E_\alpha|\psi\rangle$. Then, the inverse participation ratio
 is defined by
 \be
 \lambda=\frac{1}{\sum_{\alpha=1}^{\cal D} |c_\alpha|^4}
 =\frac{1}{{\rm Tr}(\rho_{DE}^2)}\,,
 \ee 
 which is essentially a quantity that measures the number of energy eigenstates that contribute to the state $|\psi\rangle$. 
 Note that  $1\leq \lambda\leq {\cal D}$. In fact,
 when only one energy eigenstate contributes to the state the 
 inverse participation number 
 is one, while when all energy levels equally contribute to the 
 state it is equal to ${\cal D}$.

It is worth also noting that the inverse participation ratio may be given
in terms of the infinite time average of the wave function $\phi_0$
\be
\lambda^{-1}=\lim_{T\rightarrow \infty}\frac{1}{T}\int_0^T
|\phi_0(t)|^2 dt\,.
\ee

\begin{figure}[h!]
	\begin{center}
		\includegraphics[width=0.8\linewidth]{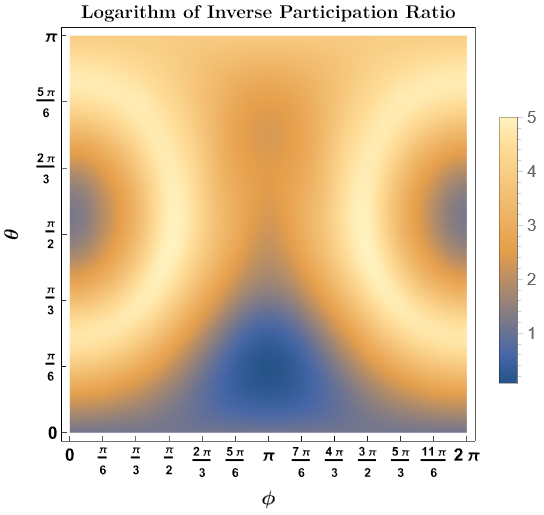}
  \end{center}
	\caption{The logarithm  of inverse participation ratio for general initial state given in 
	\eqref{initial} as a function of $\theta$ and $\phi$. The numerical computation is 
	done for $N=10$.
}
	\label{fig:IPR}
\end{figure}

Now, let us compute the inverse participation ratio for the general initial state \eqref{initial}. The results are depicted in figure \ref{fig:IPR}, which displays the logarithm of the inverse participation ratio,
$\ln \lambda$. Interestingly enough, there is a strong correlation with all the quantities we have examined thus far, including the variance of the Lanczos coefficients.

It is important to highlight that analyses of the effective inverse temperature and normalized energy suggest weak thermalization occurs for states near the edge of the energy spectrum \cite{{Banuls:2011vuw},{Sun:2020ybj},{Chen:2021},{Lin:2016egw}}. In our investigation, we found a correlation between whether a state exhibits weak or strong thermalization and its inverse participation ratio, as indicated in \cite{Prazeres:2023hce}. Specifically, the nature of a state’s thermalization is closely linked to the number of energy eigenstates contributing to it; states comprised of more energy eigenstates tend to display stronger thermalization. Our computations of the expectation values of local operators further confirm this behavior.

 Moreover, the behavior of the (log) inverse participation 
 ratio aligns perfectly with the infinite time average of complexity. This indicates that the saturation value of complexity is higher for states composed of a greater number of energy eigenstates.\cite{{Scialchi:2023bmw},{Rabinovici:2021qqt}}.

\begin{figure}[h!]
	\begin{center}
		\includegraphics[width=0.8\linewidth]{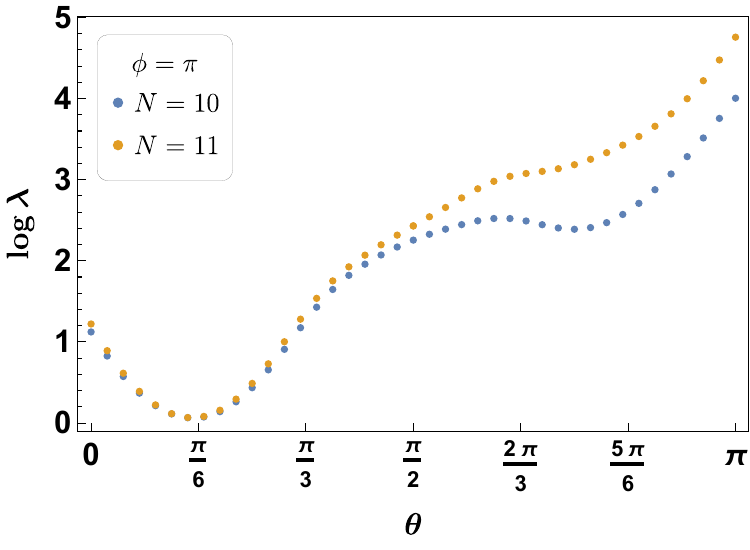}
  \end{center}
	\caption{The logarithm  of inverse participation ratio
 of $\phi=\pi$ slice for $N=10,11$. One observes that the second peak (minimum) is removed as one goes to higher $N$.
}
	\label{fig:IPR-phi=pi}
\end{figure}

It is intriguing to observe that when examining the 
$\phi = \pi$ slice, we see a behavior similar to the
variance of the Lanczos coefficients. Specifically, there are two distinct minima located around $\frac{\pi}{6}$ and  
$\frac{5\pi}{6}$. However, it is important to note that the first minimum corresponds to the state with the weakest thermalization, while the second minimum is an artifact of the finite $N$ effect. 
To illustrate this, we present the logarithm of the inverse participation ratio for $N = 10$ 
and $N = 11$ in figure \ref{fig:IPR-phi=pi}, which clearly shows that the second minimum disappears as we increase $N$. Interestingly, the corrections associated with larger 
$N$ do not significantly alter the other characteristics of the inverse participation ratio.

To conclude we note that  for $N = 11$, the behavior of the inverse participation ratio at $\phi = \pi$ closely resembles that of the energy density, with only one true minimum present; the second minimum is not genuine (see figure \ref{fig:E-phi=pi}).

\subsection{Time dependent of expectation value}

So far, we have investigated several measures to probe the nature of thermalization. It is important to note that our understanding of which states exhibit strong or weak thermalization is derived from energy behavior, or equivalently, from the infinite time average of complexity and the inverse participation ratio, as shown in the figures
\ref{fig:Ising1}, \ref{fig:complexity}
and \ref{fig:IPR}, respectively.

From these figures, we observe that the state associated with $\theta=0$ (arbitrary $\phi$), corresponding to $|Z+\rangle$, exhibits weak thermalization. In contrast, the state at 
$\theta=\pi$ (arbitrary $\phi$), corresponding to $|Z-\rangle$, demonstrates strong thermalization. As $\theta$ transitions from 0 to $\pi$, a non-trivial behavior emerges, which is evident in the 
aforementioned figures.

To validate this expectation, we can compute the expectation value of a typical operator across different states to determine if they exhibit strong or weak thermalization.
In fact, to get a better understanding of what actually happens in different points in the $\theta-\phi$ plane (initial states),  we will consider the  magnetization in the $z$ direction and compute the following quantity
\footnote{
 We have also computed the expectation value for the
magnetization in the $x$ direction,
$S_x$, in which we have found that the conclusion is the same as that 
of $S_z$ that is explicitly presented in what follows.  }
\be
\langle S_z(t)\rangle=\langle \theta,\phi| e^{-iHt}\sum_{i=1}^N\sigma^z_i e^{iHt}|\theta,\phi\rangle\,,
\ee 
for different values of  $\theta$ and $\phi$.
In examining the behavior of this expectation value, we note distinct characteristics for strong and weak thermalization.
 In the strong case, we observe a swift relaxation characterized by a quick rise or fall of the expectation value, followed by a saturation phase around the thermal value, with minor fluctuations. Conversely, weak thermalization shows oscillatory behavior from the outset, oscillating around the thermal value.

To quantify these behaviors, we calculate the ratio of the oscillation size (variance of the oscillation) to the amplitude of the first peak or trough following relaxation. This ratio serves as an indicator of thermalization strength: for states exhibiting weak thermalization, the ratio approaches one, whereas for states with strong thermalization, it is significantly less than one. By assessing this ratio for a specific initial state, we can determine its tendency toward either strong or weak thermalization. For instance, for the state $|y+\rangle$, which is expected to exhibit strong thermalization, the ratio is 
about 0.01, while for the states $|Z+\rangle$ and $|Z+\rangle$, it is about 0.5 and 0.7, respectively .

\begin{figure}
	\begin{center}
		\includegraphics[width=0.7\linewidth]{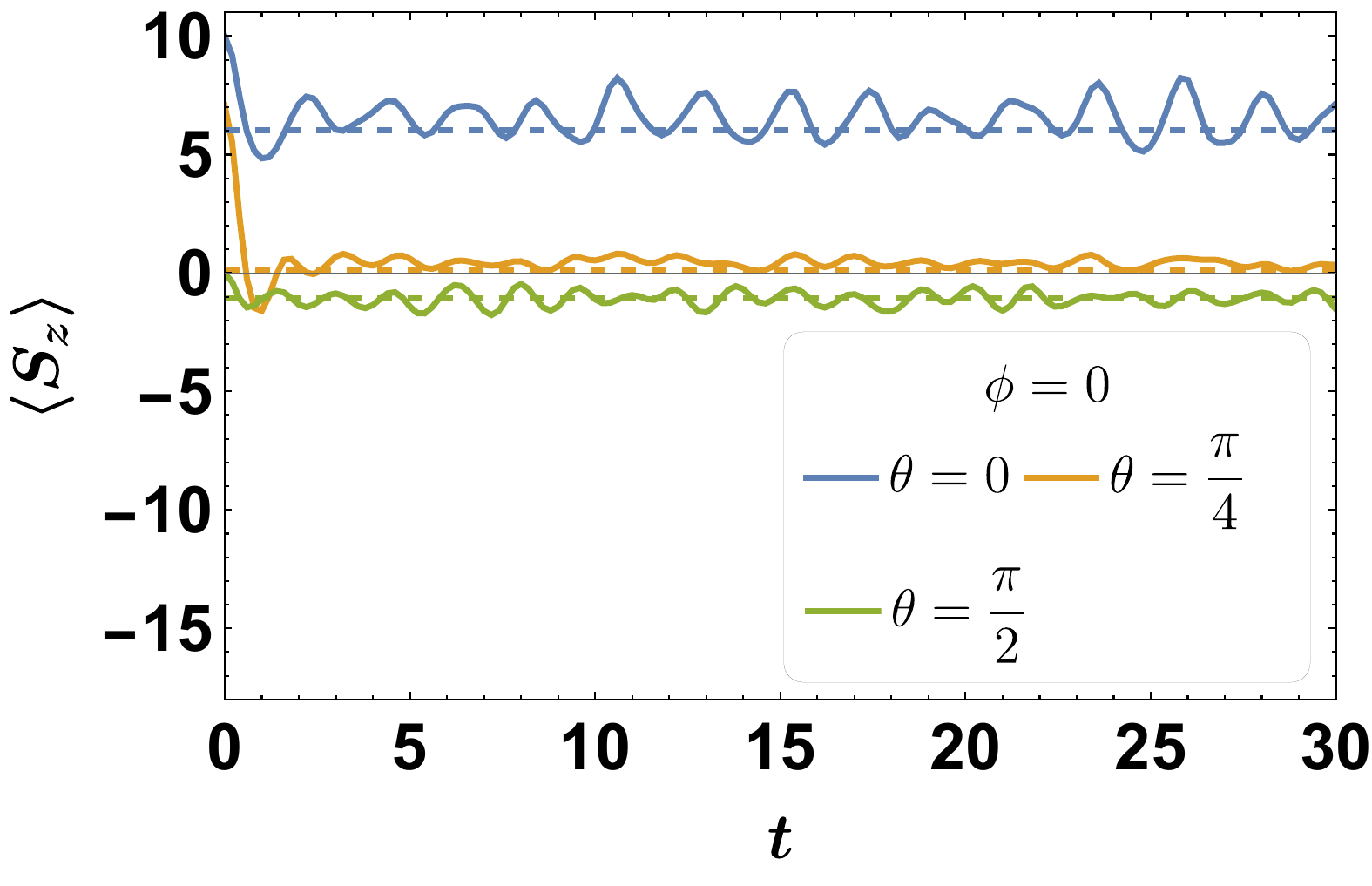}
		\includegraphics[width=0.7\linewidth]{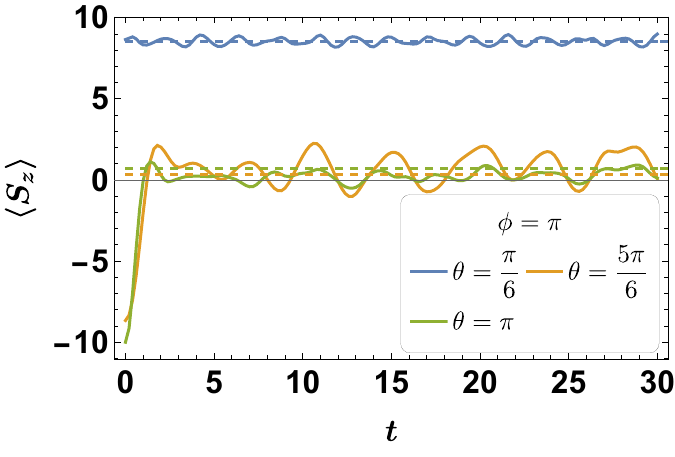}
  \end{center}
	\caption{Expectation value of $S_z$ as a function of 
	time for different initial states. As we see the weakest thermalization mostly occurs for states whose theta angle is 
 near zero while beside the ring of zero $\beta$ the strong thermalization occurs for $\theta\approx \pi$. 
}
	\label{fig:EV-sigmaz}
\end{figure}

By exploring various initial states, we have found  perfect agreement  with our expectations based on the behavior of energy. Actually, we have computed the corresponding expectation value for 
441 initial states\footnote{Since the pattern in the figure 
\ref{fig:var} is symmetric under $\phi\rightarrow 2\pi-\phi$,
we have only considered initial states located 
in $0\leq \phi\leq \pi $.} and only
few of them have been shown in this figure
\ref{fig:EV-sigmaz} which are for particular slices given by $\phi=0,\pi$. 
The 
results are relatively compatible with what suggested 
by the variance of Lanczos coefficients and in exact agreement with what is suggested by the infinite time average of complexity.

To further explore this point  let us look at $\phi=0$ slice  where 
we have presented results for
different values of $\theta$ 
in figure \ref{fig:EV-sigmaz}. By making use of these results we find weak thermalization at $\theta=0$, strong thermalization around $\theta=\pi/4$, weak thermalization again at $\theta=\pi/2$, and finally strong thermalization as we approach $\theta=\pi$.
Interestingly enough, we have found 
that the behavior is consistent with the behavior of the 
infinite time average of complexity.




Looking at $\phi=\pi$ slice, from the expectation value of $S_z$ we 
find that the weakest thermalization occurs at $\theta\approx 
\frac{\pi}{6}$ while it becomes relatively  stronger  as we move 
towards $\theta=\pi$. 
Actually evaluating the energy expectation value, one can see that the initial state $|\frac{\pi}{6},\pi\rangle$ is very close 
to an eigenstate of the Hamiltonian. Thus being localized in 
energy eigenstates one
observes an oscillatory behavior for typical operators. This can also be
seen from the inverse participation number in which for this state one
has $\lambda\approx 1$.

It is worth  also noting that  
we did not observe any 
further special point in this slice
in agreement with 
the behavior of the infinite time average of complexity and
in contrast to the 
behavior suggested by the variance in which we would expect to 
have a state with relatively weaker thermalization 
around  $\theta\approx \frac{5\pi}{6}$.

\section{Discussions}

In this paper, we have studied thermalization for a closed 
quantum system using the Krylov basis. Actually, our main motivation 
to do so is that, by definition, under time evolution 
a quantum state propagates over a subspace of Hilbert space known
as Krylov space. An advantage (at least theoretically) of working in this space is that we will have to deal with a space whose dimension is usually
smaller than the dimension of the full Hilbert space. 

In the traditional approach to quantum thermalization one usually 
has to study the expectation value of operators in the energy eigenstates.
It is believed that for a chaotic quantum system the thermalization 
occurs in the level of eigenstates that mathematically reflected 
in the statement of ETH. 

On the other hand, working within the context of Krylov space, one 
will have to compute matrix elements of local operators in the 
Krylov basis. It is then natural to expect that a similar 
concept may also show up in this context. Indeed, by making use of 
an explicit example we have shown that the matrix elements of 
local operators satisfy a condition analogous to  that of
ETH. More precisely,  
We have demonstrated that for thermalization to occur, the matrix representation of typical local operators in the Krylov basis should exhibit a specific tridiagonal form with all other elements in the matrix being exponentially small. 

We have also studied the nature of thermalization in this framework by introducing certain metrics to probe whether a given initial state exhibits weak or strong thermalization. To do so, We have observed that the nature of thermalization depends on two crucial factors: the system's Hamiltonian and the initial state. The Krylov basis and Lanczos coefficients, by construction, contain information about these elements, making them capable for studying the process of thermalization.

We have shown that the infinite time average of Krylov
complexity could provide a measure to probe the nature 
of thermalization\footnote{We also note that a certain state dependence of Krylov complexity has been studied 
in \cite{Kundu:2023hbk}} which is  in 
perfect agreement with the behavior of the effective 
inverse temperature and the density of energy. In particular,
an initial state exhibiting strong thermalization has relatively larger value for complexity saturation.

We have also suggested that the variance of Lanczos coefficients could probe the nature of thermalization to see whether for a given 
initial state the thermalization is weak or strong, too. We have seen that a state with relatively smaller (greater) variance for Lanczos coefficients $a_n$ ($b_n$) exhibits strong (weak)  thermalization. Of course, there is some mismatch between
the variance of Lanczos coefficients and the other quantities 
we have evaluated. We believe that this miss match might be 
due to the finite $N$ effect, though to explicitly show it 
we need to go to sufficiently higher $N$ and perform 
our numerical computations with extremely high precision which is out of our computational abilities. We leave exploring this point for further study.

To further explore the thermalization properties of the model
under consideration we have also evaluated the inverse participation
ratio for general initial states. We have observed that strong
thermalization occurs for states with relatively greater inverse participation ratio. In other words, a state consisting of more 
energy eigenstates is more likely to exhibit stronger thermalization. We have seen that there is a 
correlation between the behaviors of the infinite time average 
of complexity and the inverse participation ratio.

To verify our proposal we have also computed time dependence of the expectation value of local operators to explicitly probe the nature of thermalization 
for the generic initial state given by \eqref{initial}. The 
results, indeed, confirm our observation based on the behaviors 
of the infinite time average complexity, the inverse participation ratio and the variance of Lanczos coefficients. 

An interesting question we have been trying to address rather implicitly 
in this paper was the robustness of the quantities we have studied in this 
paper against the size of the system. In most numerical computations 
we have done in this paper we have set $N=10$, while in the literature the computations are done for $N=14$. It is then natural 
to see how robust the results are.

Actually, among all the quantities we have considered in this paper, 
we have presented an exact analytic expression for 
expectation value of energy which may be treated as a gauge to validate other results. 

It is clear from the exact analytic expression that 
for large $N$ limit the density of energy is independent of $N$, so that its behavior is universal which 
only depends on the parameters of the model $g$ and $h$.

We have also computed effective 
inverse temperature for $N=7$ and, surprisingly, one observes 
that it perfectly agrees with the density of energy even for large $N$, showing 
that the behavior of $\beta$ is robust against the size of the system.
Of course, as we have already mentioned the actual value
of $\beta$ is changed,
though in comparison with the numerical results available in the
litterateur for $N=14$ one finds just a few percent errors. It is 
worth emphasizing that this is also the case for other 
quantities we have studied in this paper that include
the infinite time average of complexity, the inverse 
participation number and 
 the expectation value of local operators. We note, however, that 
 for the variance of Lanczos coefficients, we expect to see significant finite $N$ effect to make it consistent with other quantities. 

To explore our idea about thermalization in the Krylov basis we have 
considered an Ising model whose Hamiltonian is given by \eqref{Ising}.
We note, however, that there is another model which has been
extensively studied in the literature whose Hamiltonian is
\be\label{xy}
H=\sum_{i=1}^{N-1}\sigma^x_i\sigma^x_{i+1}+\sigma^y_i\sigma^y_{i+1}+g\sum_{i=1}^N \sigma^y_i\,.
\ee
For the general initial state \eqref{initial} the expectation value of the
energy is 
\be\label{Exy}
E=\sin\theta\left((N-1)\sin\theta+ N g \sin\phi\right)\,.
\ee
One may also explore thermalization and its nature for this model by
evaluating  different quantities such as  effective inverse 
temperature 
and infinite time average of complexity. Doing so, one can see that 
the results are consistent with the behavior of the expectation  of 
the energy \eqref{Exy},
 that also confirms our expectations. We note that the inverse 
 participation ratio for this model has been studied in \cite{Prazeres:2023hce}.


In this paper, we have studied the 
infinite time average of the Krylov complexity and the variance of Lanczos 
coefficients associated with the spread of an initial state 
\cite{Balasubramanian:2022tpr}. We note, however, that 
the same question as studied in this paper can be also addressed 
using Lanczos coefficients associated with operator growth \cite{Parker:2018yvk} \footnote{We note that operator and
state growth may be studied within a universal framework\cite{Alishahiha:2022anw}}. Essentially in our context, it corresponds 
to changing the picture from Schr\"odinger to Heisenberg.

In the Heisenberg picture of quantum mechanics, we are dealing with the
operators and the time evolution is attributed to the operator 
\be
{\cal O}(t)=e^{-iHt}{\cal O} e^{iHt}\,.
\ee
Defining an inner product in the  space of operators acting on the 
Hilbert space, one can construct the Krylov basis for the operator 
starting with an initial operator ${\cal O}$. The first element 
is identified with the initial operator $O_0={\cal O}$ (which we assume to be normalized with respect to the inner product)
and the other elements  may be constructed recursively as follows
\be
{\hat O}_{n+1}={\cal L} O_{n}-\hat{b}_n O_{n-1},\;\;\;O_n=\hat{b}_n^{-1} \hat{O}_n,
\ee
where ${\cal L}O_n=[H,O_n]$ and $\hat{b}^2_n=|\hat{O}_n\cdot 
\hat{O}_n|$ is Lanczos coefficients. The procedure stops for $n={\cal D}_O\leq {\cal D}^2-{\cal D}+1$  \cite{Rabinovici:2020ryf} that is the dimension 
of Krylov space for the operator.
Here we denote the Lanczos coefficients with a hat
to avoid confusion with those  defined in the  
Krylov basis for state in \eqref{LC-state}. Using this basis one
has
 \be
 {\cal O}=\sum_{n=1}^{{\cal D}_O-1}\,i^n\,\varphi_n(t)\, O_n\,.
 \ee
  Note that with this notation $\varphi_n(t)$ is real and satisfies the following equation
 \be
 \partial_t\varphi_n(t)=\hat{b}_{n} \varphi_{n-1}-\hat{b}_{n+1}\varphi_{n+1}\,.
 \ee
 In this context, we could also look for the variance of Lanczos coefficients in the operator 
 picture. To study Lanczos coefficients for the Ising model \eqref{Ising} we may consider a
 generic initial operator as follows
 \bea
 {\cal O}_{\theta,\phi}=\prod_{i=1}^N {\cal O}_i(\theta,\phi)\,,\\ \nonumber
 \eea
  where ${\cal O}_i$ is defined in \eqref{OPi}. It is worth noting that since this 
 initial state \eqref{initial} is the eigenstate  of the above
 operator, in this case, we are essentially studying the time evolution of  
 density matrix associated with the initial state
 $
 {\cal O}_{\theta,\phi}=
\rho(\theta,\phi)=|\theta,\phi\rangle\langle \theta,\phi|$. 

One can also study the variance of Lanczos coefficients $\hat{b}_n$ associated 
with the initial density matrix. Doing so, one finds the corresponding variance results in the same conclusion as that 
for the state studied in the previous section. An  interesting observation  we have made is that the behavior of variance 
$\hat{b}_n$ in operator growth is actually identical
with that obtained from $a_n$ in state growth. It would be interesting to understand this point better.

 \section*{acknowledgements}
 
 We would like to thank 
 Ali Mollabashi, Mohammad Reza Mohammadi Mozaffar,  Mohammad Reza Tanhayi and Hamed Zolfi for useful discussions. We would also like to thank the School of Physics of the Institute for Research in Fundamental Sciences (IPM) for providing computational facilities.
 M.A. Would also like to thank
 Souvik Banerjee for discussions on different aspects of Krylov space. 
 Some numerical computations related to this work were carried out at IPM Turin Cloud Services \cite{turin}. This work is based upon research founded by Iran National Science Foundation (INSF) under project No 4023620.



{}


\end{document}